\documentclass[aps,prb,twocolumn,superscriptaddress,print]{revtex4}
\usepackage{amssymb,amsmath}
\usepackage{graphicx}
\usepackage{multirow}
\usepackage{hyperref}
\usepackage{appendix}
\usepackage{mathtools}
\usepackage{amsmath}
\usepackage{verbatim}
\usepackage{gensymb}
\usepackage{subfigure}
\usepackage{bm}
\usepackage{threeparttable}
\usepackage{booktabs}
\usepackage{braket}
\usepackage{xcolor}
\usepackage[normalem]{ulem}

\begin{document}
\title{Pressure and electric field dependence of quasicrystalline electronic states in 30{\degree} twisted bilayer graphene}
\author{Guodong Yu}
\email{guodong.yu@whu.edu.cn}
\affiliation{Key Laboratory of Artificial Micro- and Nano-structures of Ministry of Education and School of Physics and Technology, Wuhan University, Wuhan 430072, China}
\affiliation{Institute for Molecules and Materials, Radboud University, Heijendaalseweg 135, NL-6525 AJ Nijmegen, Netherlands}
\author{Mikhail I. Katsnelson}
\affiliation{Institute for Molecules and Materials, Radboud University, Heijendaalseweg 135, NL-6525 AJ Nijmegen, Netherlands}
\author{Shengjun Yuan}
\email{s.yuan@whu.edu.cn}
\affiliation{Key Laboratory of Artificial Micro- and Nano-structures of Ministry of Education and School of Physics and Technology, Wuhan University, Wuhan 430072, China}
\affiliation{Institute for Molecules and Materials, Radboud University, Heijendaalseweg 135, NL-6525 AJ Nijmegen, Netherlands}

\begin{abstract}
30{\degree} twisted bilayer graphene demonstrates the quasicrystalline electronic states with 12-fold symmetry. These states are however far away from the Fermi level, which makes conventional Dirac fermion behavior dominating the low energy spectrum in this system. By using tight-binding approximation, we study the effect of external pressure and electric field on the quasicrystalline electronic states. Our results show that by applying the pressure perpendicular to graphene plane one can push the quasicrystalline electronic states towards the Fermi level. Then, the electron or hole doping of the order of $\sim$ $4\times10^{14}$ $cm^{-2}$ is sufficient for the coincidence of the Fermi level with these quasicrystalline states. Moreover, our study indicates that applying the electric field perpendicular to the graphene plane can destroy the 12-fold symmetry of these states and break the energy degeneracy of the 12-wave states, and it is easier to reach this in the conduction band than in the valence band. Importantly, the application of the pressure can recover the 12-fold symmetry of these states to some extent against the electric field. We propose a hybridization picture which can explain all these phenomena.
\end{abstract}

\maketitle
\section{Introduction}
The linear band structure of graphene can be modified efficiently by stacking one layer onto another. $AA$ stacked bilayer graphene shows the band structure of two shifted Dirac cones above and below the Fermi level.\cite{BG_AA0} The bilayer in AB stacking is characterized (in the simplest approximation)
by a band structure with parabolic touching point.\cite{AB_orig0,AB_orig1,BG_AB0,BG_AB1,BG_AB2} Moreover, a twist angle between two layers offers an additional degree of freedom to tune the electronic properties. For example, the slightly twisted bilayer graphene at the magic angle as a model system of strongly correlated electrons has drawn much attention due to the novel electronic properties, such as the flat band\cite{TBG_flatband,TBG_flatband1}, unconventional superconductivity\cite{BG_superconducting,TBG_supertivity1,TBG_supertivity2}, correlated insulator phases\cite{TBG_insulator_phase}, etc. Besides, if the twist angle $\theta$ does not satisfy the commensurate condition\cite{TBG_commensurate_condition}, namely $\cos{\theta}={{n^2+4nm+m^2}\over{2(n^2+nm+m^2)}}$, where $n$ and $m$ are integers, the corresponding bilayer structures will not posses the translational symmetry. Falling into this classification, the 12-fold symmetry and the quasi-periodicity of the 30{\degree} twisted bilayer graphene has been demonstrated by various measurements, such as the Raman spectroscopy, low-energy electron microscopy/diffraction (LEEM/LEED), transmission electron microscopy (TEM) and scanning tunneling microscopy (STM) measurements.\cite{pnas_QC,science_QC,cm_QC,30TBG_STM,30TBG_SiC_arXiv}.

By now, the 30{\degree} twisted bilayer graphene has been grown successfully on some substrates, such as SiC\cite{science_QC,30TBG_SiC_arXiv}, Pt\cite{pnas_QC}, Cu-Ni\cite{cm_QC} and Cu\cite{30tBG_on_Cu_ACSNano, 30tBG_onCu_arXiv} surfaces. This emergent quasicrystal consisting of two graphene sheets with perfect crystalline has attracted increasing attention because of the coexistence of the quasicrystalline nature and the relativistic properties.\cite{30TBG_STM,30TBG_SiC_arXiv,30TBG_Ultrafast_unbalance_arXiv,30TBG_localization,30TBG_quantum_oscillation,science_QC,pnas_QC,cm_QC,30TBG_superlubricity}. Angle-resolved photoemission spectroscopy (ARPES) measurements indicated that the interlayer interaction between the two layers leads to the emergence of the mirror-symmetric Dirac cones inside the Brillouin zone of each graphene layer\cite{science_QC,pnas_QC,30TBG_SiC_arXiv} and a gap opening at the zone boundary\cite{pnas_QC}. The critical eigenstates\cite{30tbg_Moon,30TBG_localization} and quantum oscillations\cite{30TBG_quantum_oscillation} were predicted theoretically due to the quasi-periodicity. 

As an extrinsic quasicrystal consisting of two graphene crystalline layers, the appearance of the 12-fold symmetric electronic states originates from the weak Van der Waals interaction between the two layers. The strength of the interlayer coupling determined by the interlayer spacing affect the quasicrystalline electronic states directly. To apply the external pressure is an efficient method to control the interlayer spacing, especially for weakly interacting materials. It was shown theoretically that the external pressure perpendicular to graphene plane can enhance the interlayer coupling and modify the magic angle value and associated density of states of slightly twisted bilayer graphene.\cite{TB_wannier_pressure} Besides the interlayer interaction, the interaction among the degenerated 12 waves in the reciprocal space is the direct reason to form the 12-fold symmetric electronic states\cite{30tbg_Moon}. Breaking the degeneration of the 12 waves should modify the 12-fold symmetry of these states obviously. As an example, the electric field perpendicular to the graphene plane is one way to break the degeneration, which differ the energy of the 6 waves in one layer from the 6 waves in another layer.   

In this paper, our purpose is to study the dependence of the quasicrystalline electronic states on external pressure and electric field perpendicular to graphene plane in 30{\degree} twisted bilayer graphene. Although the 12-fold symmetric electronic states form in 30{\degree} twisted bilayer graphene, they do not contribute to the most of electronic properties because of their large distance from the Fermi level. In this study, we find out the way to tune the energies of these 12-fold symmetric electronic states closer to Fermi level and discuss the stability of these 12-fold symmetric states under the external pressure and electric field. 

\begin{figure}[!htbp]
\centering
\includegraphics[width=7 cm]{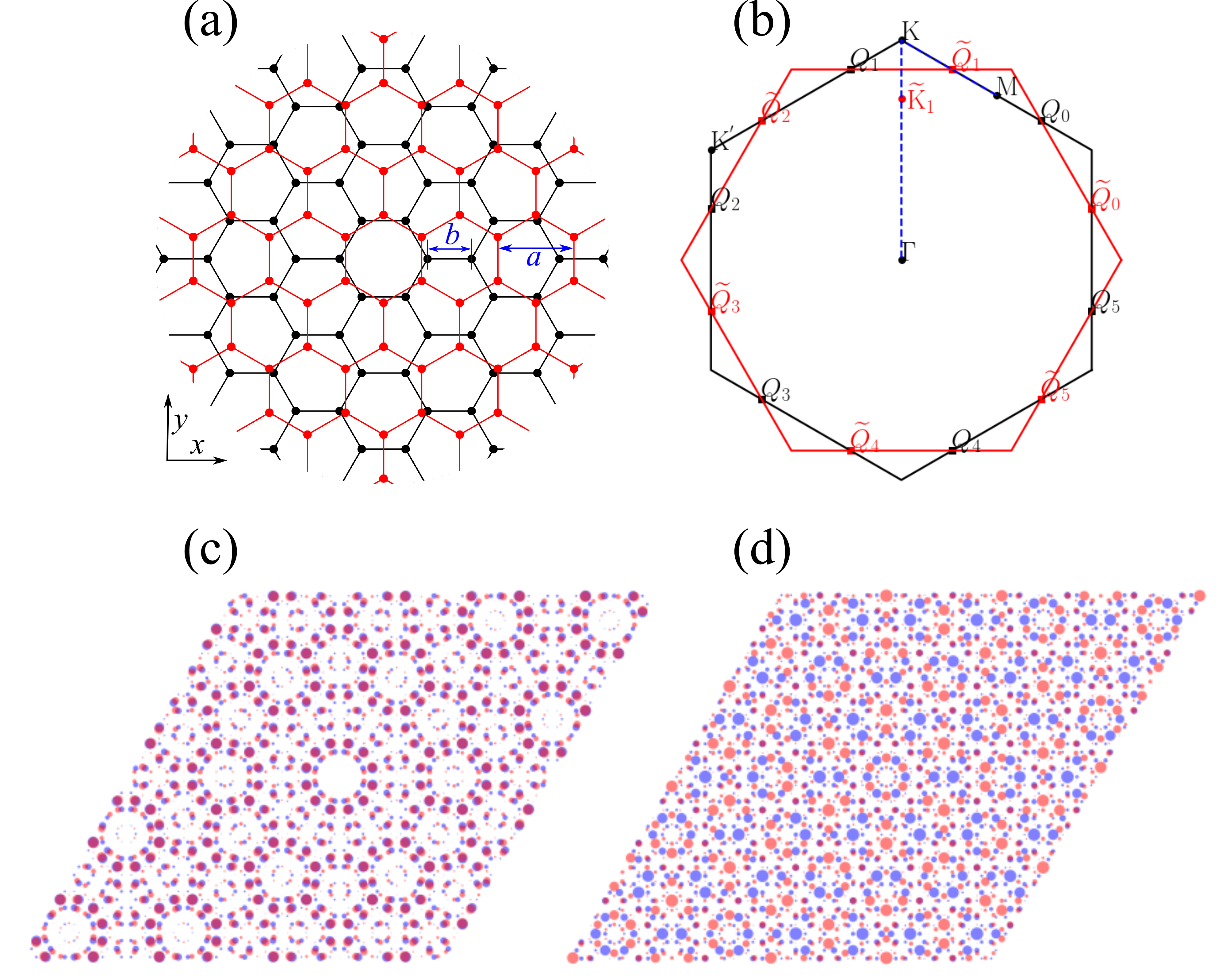}
\caption{(a) The structure of 30{\degree} twisted bilayer graphene with the periods of bottom (black) and top (red) layers being $a\;(=2.46 \; \rm{\AA})$ and $3b\;(=\sqrt{3}a=4.26 \; \rm{\AA})$, respectively, along $x$ axis. In the 15/26 approximant, the lattice constant $a$ of the top layer changes to be $a^{\prime}\;(=2.458 \; \rm{\AA})$. (b) The Brillouin zones of the two layers. $\widetilde{K}_1$ is the end-point after the strongest scattering process from $K^{\prime}$ point due to the interlayer interaction. $\widetilde{K}_1$ and $K$ are the mirror-symmetric points with respect to the mirror-line $Q_1-\widetilde{Q}_1$. The points $Q_{i}, i=0,...,5$ of bottom layer and the points $\widetilde{Q}_{i}, i=0,...,5$ of top layer are degenerated in energy, the hybridization of which, namely 12-wave Hamiltonian at $k_0=0$ \cite{30tbg_Moon}, results in the emergence of the quasicrystal electronic states. (c) and (d) are the quasicrystal electronic states with 12-fold symmetry at the VBM and CBM, respectively, of any $Q_i$ or $\widetilde{Q}_i$ point, which are calculated by 15/26 approximant. Blue and red dots correspond to the occupation numbers on the bottom and top layers, respectively, with larger occupation number denoted by larger dot.}
\label{fig:struct}
\end{figure}

\section{Methods}
\subsection{Tight-binding model}
The tight-binding model based on the maximally localized Wannier function\cite{TB_wannier}, which possess the higher angular momentum components ($m=3n$ with $n\in \mathbb{Z}$) than $p_z$ orbital ($m=0$) for twisted bilayer graphene, is adopted to study the graphene quasicrystal under the finite external pressure. The intralayer hopping energies up to the eighth nearest neighbors are used to describe graphene monolayer, which are -2.8922, 0.2425, -0.2656, 0.0235, 0.0524, -0.0209, -0.0148 and -0.0211 eV, respectively, from first to eighth nearest neighbors. The interlayer hopping described by a functional form depends on both distance and orientation. After ignoring the high order terms ($|n|>2$), which are vanishingly small, the interlayer hopping function reads
\begin{equation}
\begin{split}
  t(\bm{r}) = V_0(r) +  V_3(r)[cos(3\theta_{12}) + cos(3\theta_{21})] + \\ V_6(r)[cos(6\theta_{12}) + cos(6\theta_{21})].
\end{split}
\label{interhopping}
\end{equation}  
$\bm{r}$ is the projection of the vector connecting two sites on graphene plane. $r$=$|\bm{r}|$ descries the projected distance between two Wannier functions. $\theta_{12}$ and $\theta_{21}$ are the angles between the projected interlayer bond and the in-plane nearest-neighbor bonds, which describe the relative orientation of the two Wannier functions. The three radial functions depend on ten hopping parameters:
\begin{equation}
\begin{split}
V_0(r) = \lambda_0e^{-\xi_0\bar{r}^2}cos(\kappa_0\bar{r}),\\
V_3(r) = \lambda_3\bar{r}^2 e^{-\xi_3 (\bar{r}-x_3)^2},\\
V_6(r) = \lambda_6 e^{-\xi_6 (\bar{r}-x_6)^2} sin(\kappa_6\bar{r}).
\end{split}
\end{equation}

For a twisted bilayer graphene, the relationship between the interlayer spacing compression ($\varepsilon=1-h/h_0$ with $h$ and $h_0$ being the interlayer spacings with finite and zero external pressures, respectively.). and external pressure $P$ satisfies Murnaghan equation of state \cite{Murnaghan_equation},
\begin{equation}
P = A(e^{B\varepsilon} - 1).
\end{equation} 
The parameters $A$ and $B$ were determined to be 5.73 GPa and 9.54, respectively, from previous study \cite{TB_wannier_pressure} by fitting the density functional theory results. The interlayer spacing compression dependence of the ten interlayer hopping parameters are well described by a quadratic fit
\begin{equation}
y_i(\varepsilon)=c_i^{(0)} -c_i^{(1)}\varepsilon + c_i^{(2)}\varepsilon^2,
\end{equation}   
where $y_i$ ($i=1,...,10$) stands for any one of the ten interlayer hopping parameters. The coefficients $c_i^{(0)}$, $c_i^{(1)}$ and $c_i^{(2)}$ for all interlayer hopping parameters are listed in Table \ref{table:hopping_ci}. In this paper, the external pressure less than 30 GPa are under consideration, because previous calculations did not show significant reconstruction of the graphene bilayer under pressure even up to 30 GPa,\cite{TB_wannier_pressure} but there may be a phase transition of the encapsulating hBN substrate around 9 GPa\cite{hBN_pressure}.

\begin{table}[htp!]
\caption{The ten interlayer hopping parameters (in units of eV).}
\centering
\begin{tabular}{c c c c}
\hline\hline                       
 %inserts double horizontal lines
 $y_i$ & $c_i^{(0)}$ & $c_i^{(1)}$ & $c_i^{(2)}$ \\ 
 \hline                  
 $\lambda_0$ & 0.310 & -1.882 & 7.741  \\
 $\xi_0$ & 1.750 & -1.618 & 1.848  \\
 $\kappa_0$ & 1.990  & 1.007 & 2.427  \\
 $\lambda_3$ & -0.068 & 0.399 & -1.739 \\
 $\xi_3$ & 3.286 & -0.914 & 12.011 \\ 
 $x_3$ & 0.500 & 0.322 & 0.908\\
 $\lambda_6$ & -0.008 & 0.046 & -0.183 \\
 $\xi_6$ & 2.727 & -0.721 & -4.414 \\
 $x_6$ & 1.217 & 0.027 & -0.658\\
 $\kappa_6$ & 1.562 & -0.371 & -0.134\\
 \hline
  \end{tabular}
 \label{table:hopping_ci}
 \end{table}

\begin{figure*}[!htbp]
\centering
\includegraphics[width=15 cm]{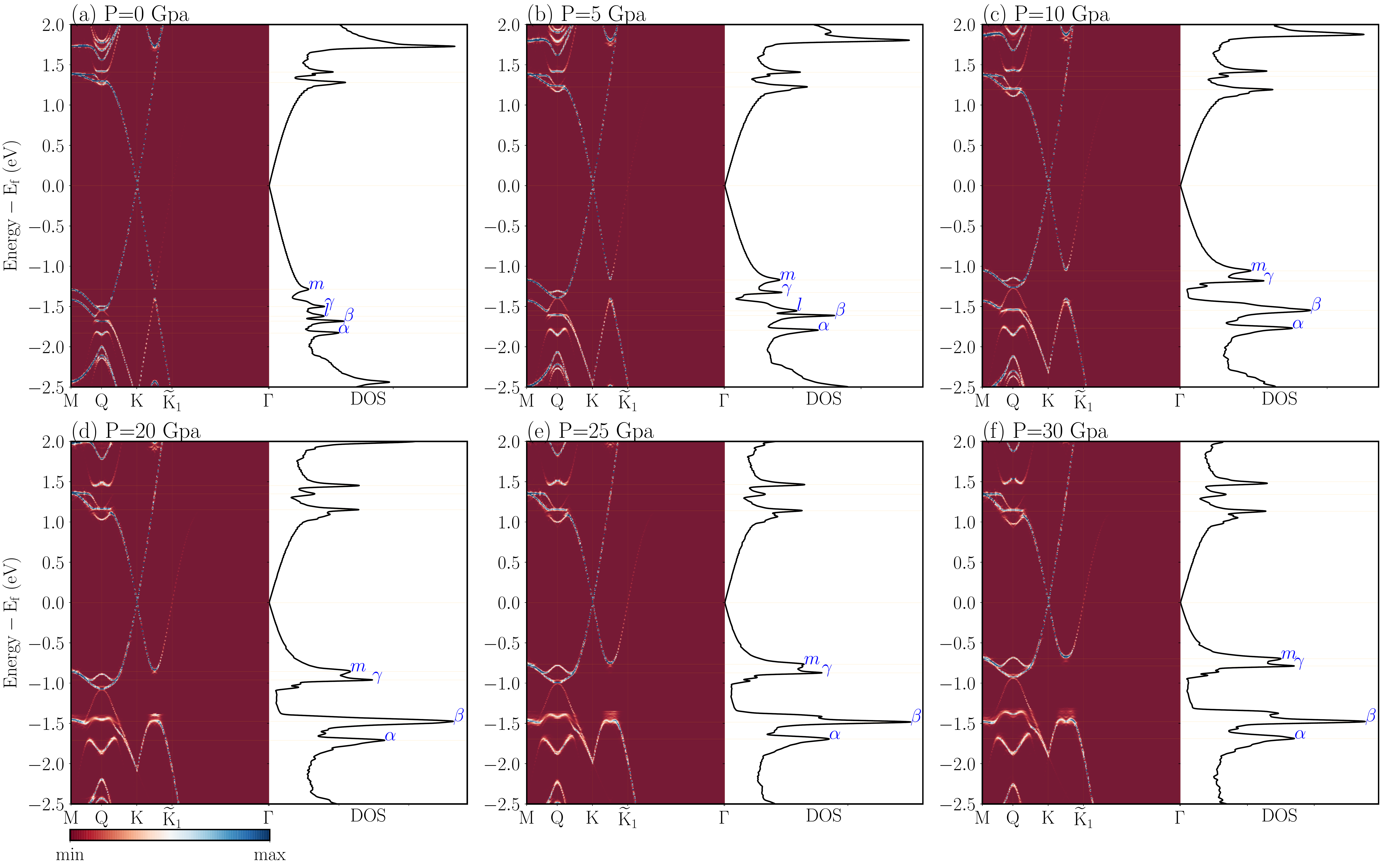}
\caption{The comparisons of effective band structures and density of states at several pressures. The k-path in the effective band structure is along the dashed blue line shown in Fig. \ref{fig:struct}(b). $Q$ is $\widetilde{Q}_1$ in this figure. The eigen-states at VBM and CBM around $Q$ point are shown in Fig. \ref{fig:struct}(c) and (d), respectively. The pressure affects the density of states in the valence band stronger than in the conduction band.}
\label{fig:ebs_dos}
\end{figure*}

\section{Results and discussion}
In order to study 30{\degree} twisted bilayer graphene under pressure in the framework of band theory, the 15/26 approximant\citep{Yu2019,tMGQ} has been chosen to calculate to calculate its electronic properties. The 15/26 approximant is a periodic Moir\'{e} pattern, which is obtained by compressing the top layer slightly with the lattice constant changing from 2.46 {\AA} to 2.458 {\AA}. Accordingly, the two layers share the commensurate period $15\times \sqrt{3}a=26\times\widetilde{a}$, where $\sqrt{3}a$ and $\widetilde{a}$ are the periods of bottom and top layers along $x$ direction, respectively. It has been proven that the 15/26 approximant can reproduce the electronic properties of the original 30{\degree} twisted bilayer graphene accurately and the 12-fold symmetry of the quasicrystalline electronic states can be distinguished within the unit cell\citep{Yu2019}. Furthermore, by unfolding the band structure of the 15/26 approximant into the premitive unit cell of graphene\cite{BandUnfold_SF,BandUnfold_EBS}, the effective band structure can be derived. It can be used to compare the ARPES measurement. 

In the case of zero pressure, the interaction between two layers in 30{\degree} twisted bilayer graphene results in the appearance of five new van Hove singularities of density of states in the valence band, three of which, labelled by $\alpha$, $\beta$ and $\gamma$, are associated with the critical wave functions \citep{30tbg_Moon}, which show the quasicrystalline nature clearly. Different from the localized states, the critical wave functions still spread in a big spatial area, so our approximant model can not reproduce the real critical wave function. However, the quasicrystal nature can be recognized by the eigen-states with 12-fold symmetry within the unit cell, because 12-fold symmetry is forbidden in crystallines with translational symmetry. In Fig. \ref{fig:struct}(c) and (d), we plot the charge distributions of VBM and CBM around $Q_0$. They show the 12-fold symmetry in the unit cell of the 15/26 approximant, namely quasicrystalline nature. Actually, all $Q_i$ and $\widetilde{Q}_i$ points given in Fig. \ref{fig:struct}(b) are degenerated in energy, and they give exact the same electronic states. Because we focus on mainly the quasicrystalline electronic states in this paper, in the following text, the VBM and CBM always stand for the valence-band minimum and conduction-band maximum at any $Q_i$ or $\widetilde{Q}_i$ point or the case $k_0=0$ in the k-space tight-binding method (see below for the details of this method). The appearance of the 12-fold symmetric states can be attributed to the degenerated 12-wave ($Q_0-Q_5$ and $\widetilde{Q}_0-\widetilde{Q}_5$) interaction. However, these quasicrystalline electronic states are far away from the Fermi level, so they contribute much weaker than Dirac electrons in electronic properties.

\begin{figure}[!htbp]
\centering
\includegraphics[width=6 cm]{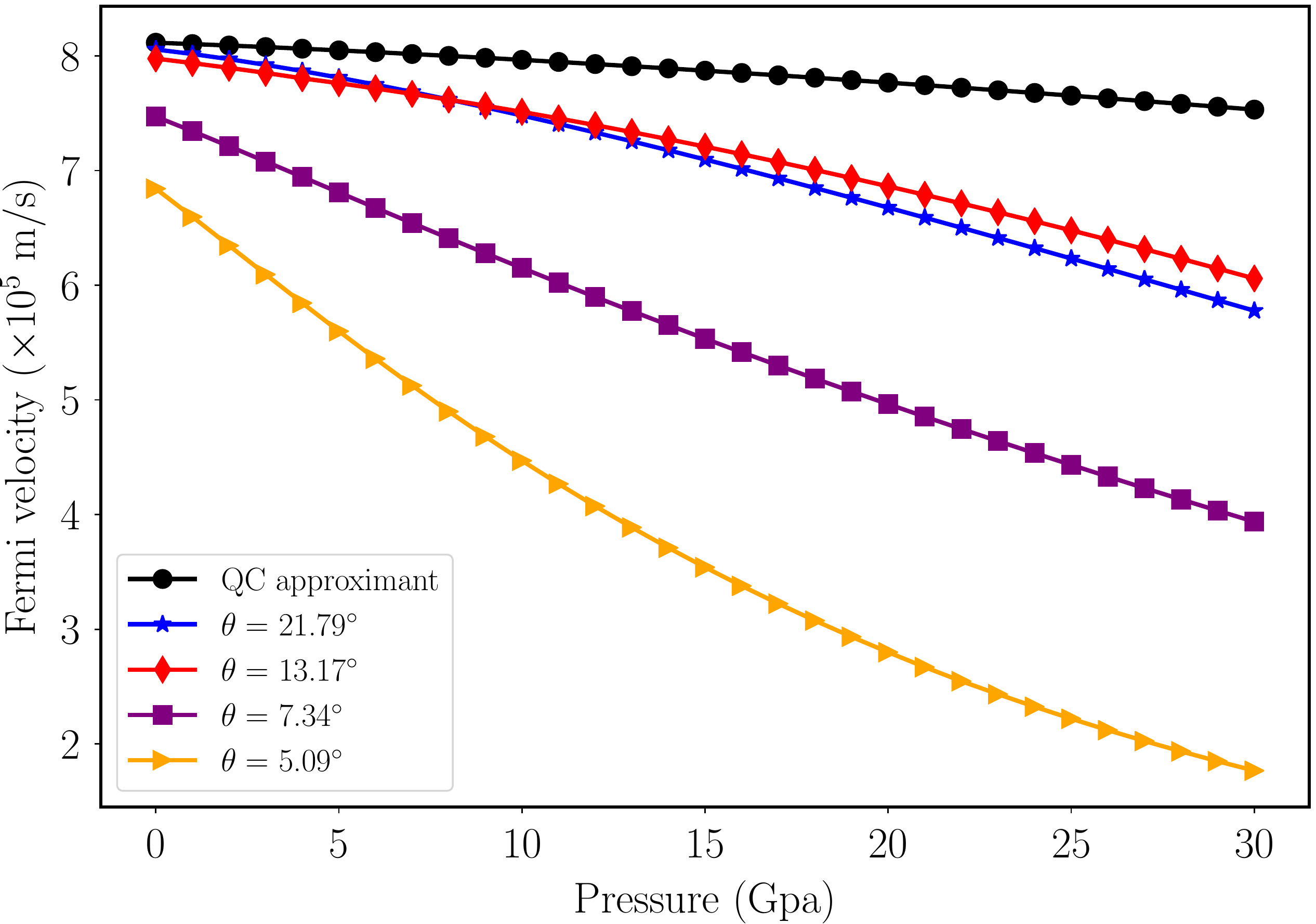}
\caption{The comparison of Fermi velocity reduction with pressure increasing for several twisted bilayer graphene at different twisted angle $\theta$. 30{\degree} twisted bilayer graphene is calculated by 15/26 approximant (QC approximant).}
\label{fig:vfs}
\end{figure}

\begin{figure}[!htbp]
\centering
\includegraphics[width=7 cm]{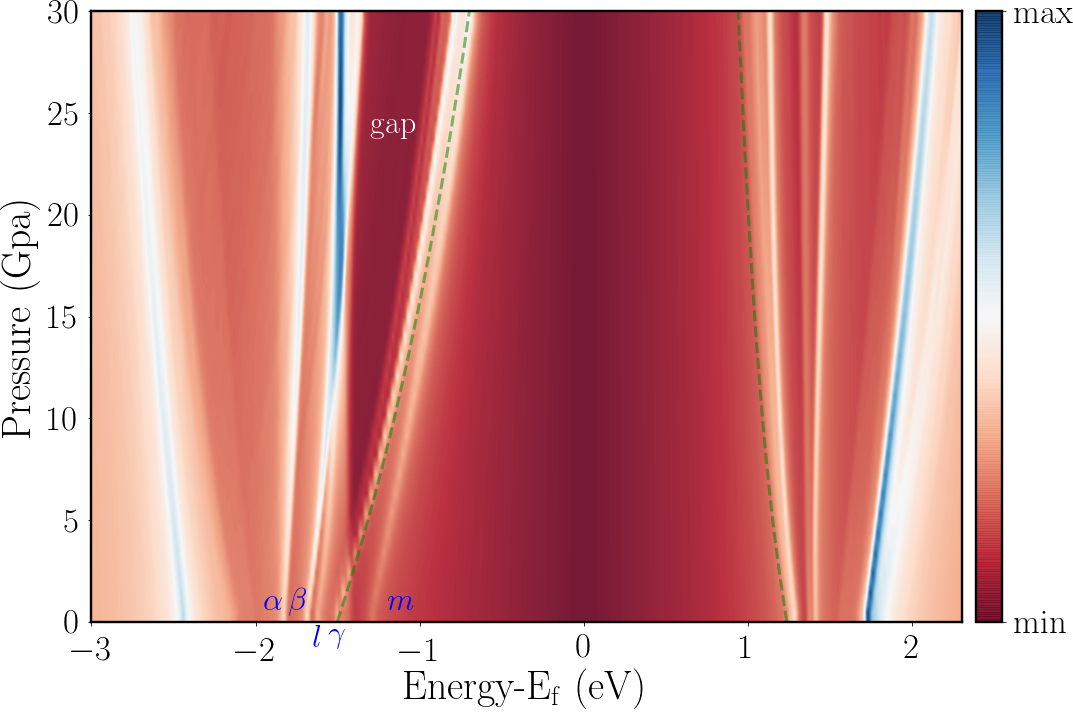}
\caption{The density of states of 30{\degree} twisted bilayer graphene under the pressure less than 30 GPa. Five peaks in the valence band are marked. $\beta$ and $l$ peaks as well as $\gamma$ and $m$ peaks merge gradually as pressure increases. The positions of the VBM and CBM at $Q$ point, namely the energies of the quasicrystal electric states, are shown by the green dashed lines.}
\label{fig:dos}
\end{figure}

\begin{figure}[!htbp]
\centering
\includegraphics[width=6 cm]{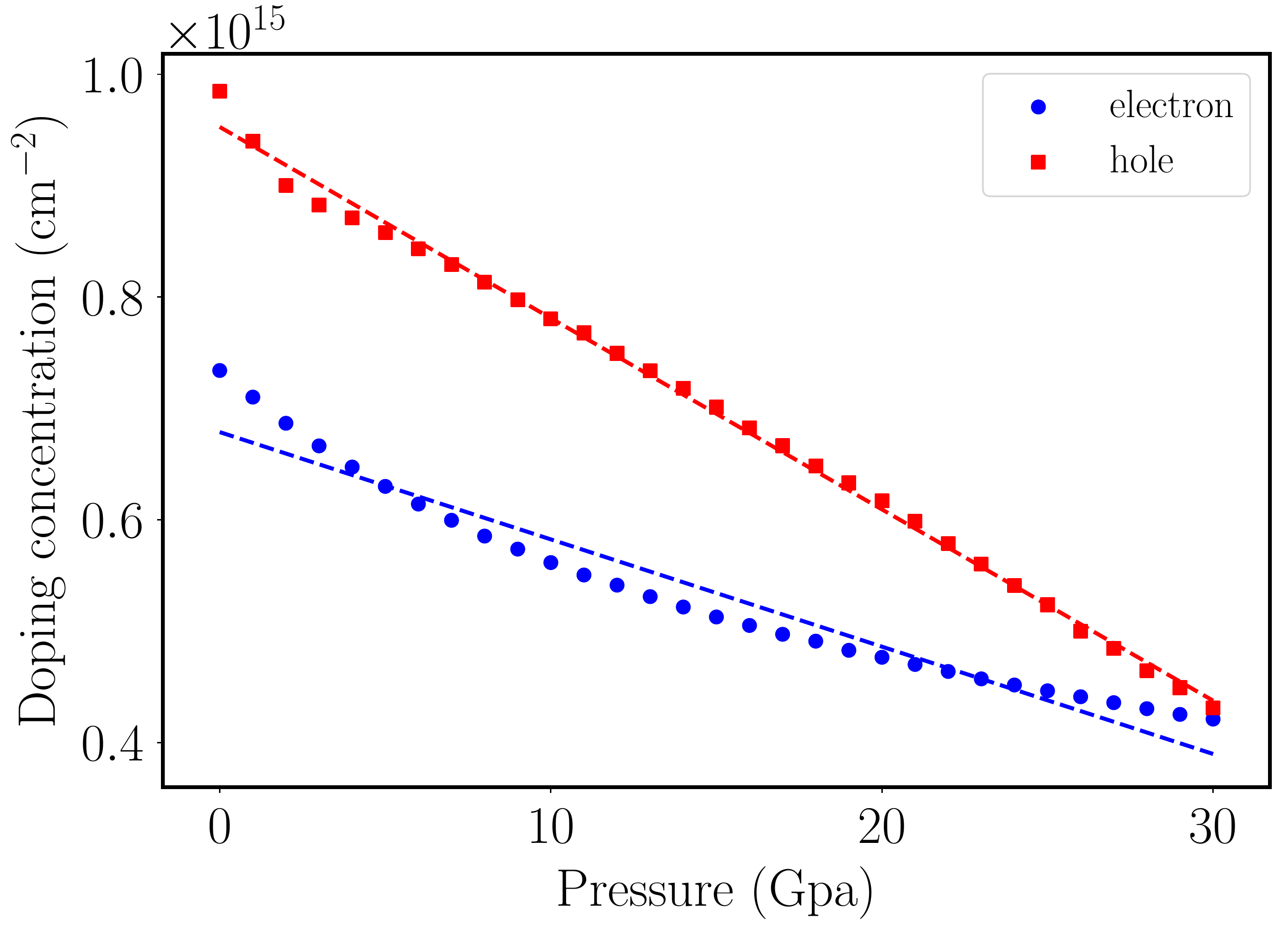}
\caption{The doping concentration of holes and electrons needed to make the Fermi level to meet the VBM and CBM of $Q$ point, respectively, for the pressure less than 30 GPa. }
\label{fig:doping}
\end{figure}

Fortunately, our results indicate that applying the external pressure to graphene quasicrystal is a valid method to push the quasicrystalline electronic states towards the Fermi level. In Fig. \ref{fig:ebs_dos}, we compare the density of states and effective band structures of 30{\degree} twisted bilayer graphene at several pressures. It can be found that the peaks $\beta$ and $m$ as well as the peaks $\gamma$ and $l$ merge gradually as the pressure increases, and the continuous evolutions of these peaks are given in Fig. \ref{fig:dos}. As the pressure increases, there is a big gap appearing between peaks $\beta$ and $\gamma$. The partial reason is the increasing interaction between the Dirac cone at $K$ and its mirror-symmetric Dirac cone at $\widetilde{K}_1$, which also reduce the Fermi velocity (shown in Fig. \ref{fig:vfs}). Comparing with some twisted bilayer graphene at smaller twist angles, the Fermi velocity in 30{\degree} twisted bilayer graphene is affected much weaker because of the largest distance between closest Dirac cones at the case of 30{\degree} twist angle. It means that the Dirac fermion behavior near the Fermi level is robust. Let us focus on the VBM and CBM now. Their 12-fold symmetry are always kept in any pressure less than 30 GPa. The reason is that as the pressure increases the interlayer coupling become stronger, but the energy degeneration of the 12 waves always remains no matter how large pressure is applied. The VBM and CBM positions are shown in Fig. \ref{fig:dos} and remarked by two green dashed lines. The results indicate that, as the pressure increases from 0 to 30 GPa, the quasicrystal electrons move gradually towards the Fermi level and their positions derivate from any peak of the density of states in energy. It means that it become easier gradually to tune the Fermi levels by electron or hole doping to enhance the contribution of the quasicrystalline electronic states in electronic properties as the pressure increases. We will explain this phenomenon below. In Fig, \ref{fig:doping}, we show the doping concentration of electrons and holes that are needed to tune the Fermi levels at the CBM and VBM. For two dimensional materials, the magnitude of the doping concentration $\sim10^{14}\; cm^{-2}$ can be realized easily by using ionic liquid gates.\cite{doping0,doping1}

\begin{figure*}[!htbp]
\centering
\includegraphics[width=14 cm]{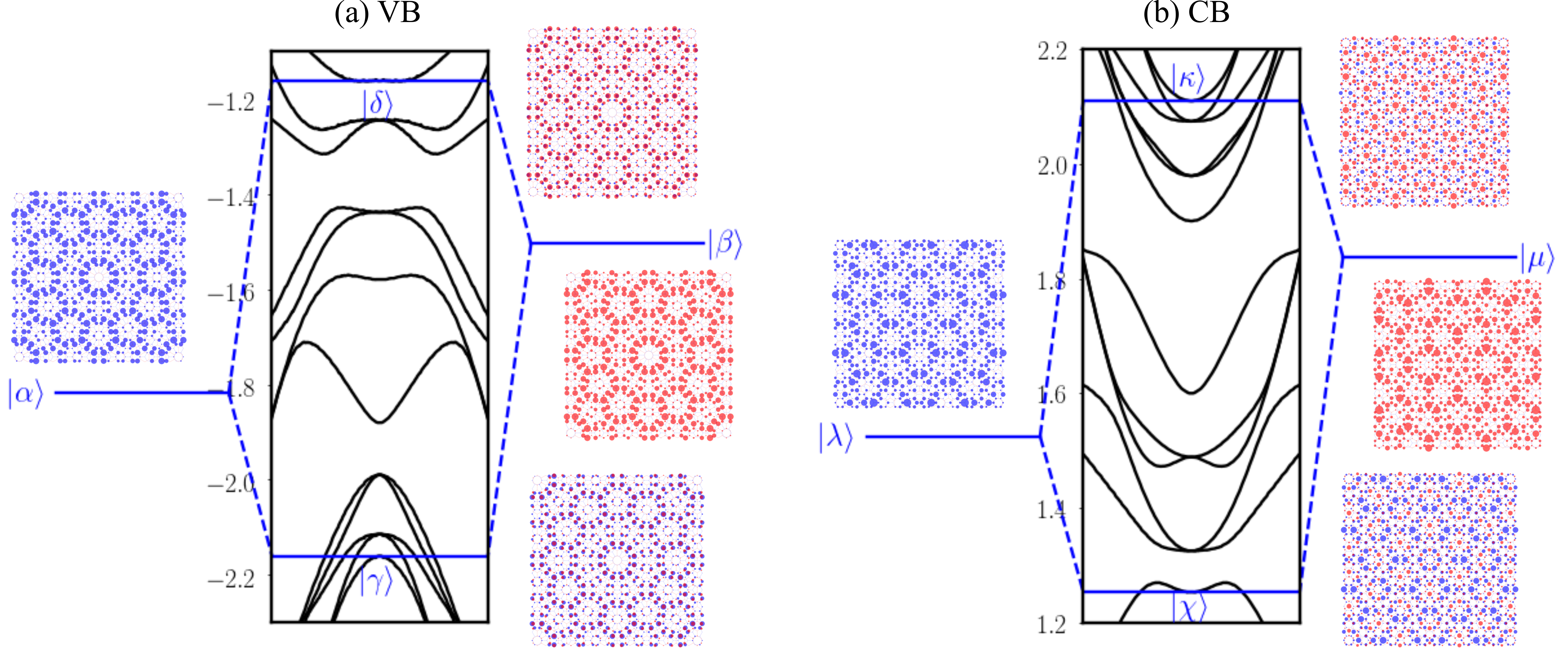}
\caption{The quasi-band structures around $k_0=0$ calculated from 12-wave Hamiltonian and the hybridization pictures for constructing the quasicrystal electronic states in the valence band (a) and the conduction band (b), respectively. The picture shows the example at the 0.1 eV/{\AA} electric field and 5 GPa pressure. It clearly shows that the electric field destroys the 12-fold symmetry of the quasicrystalline electronic states.}
\label{fig:hybrid}
\end{figure*}

\begin{figure*}[!htbp]
\centering
\includegraphics[width=14 cm]{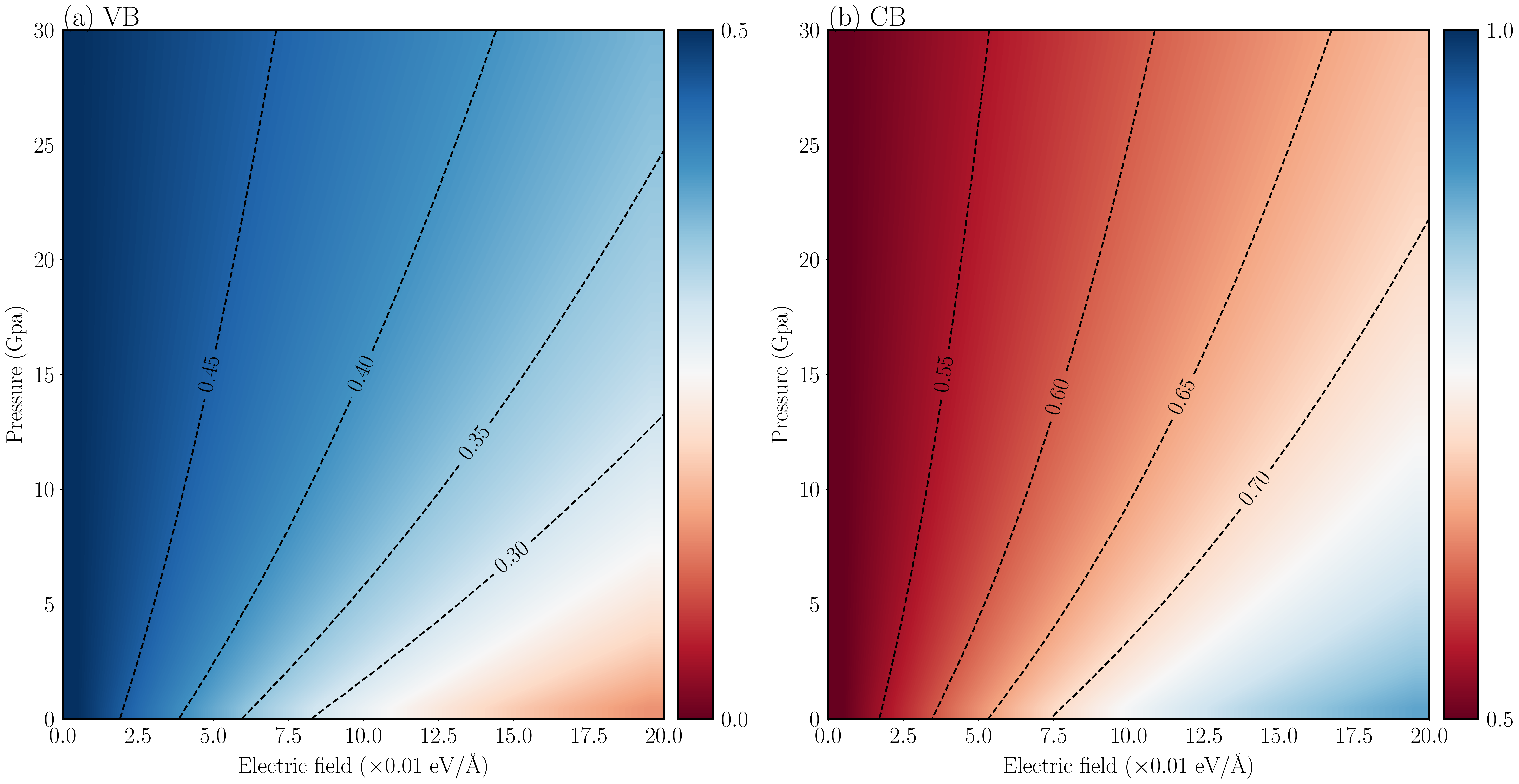}
\caption{The occupation numbers on the bottom layer of the functions $\left|\delta\right>$ $\left| C_{\gamma\alpha}\right|^2$ (a) and $\left| \chi \right>$ $\left|C_{\chi\lambda} \right|^2$ (b), respectively, under the pressure less than 30 GPa and electric field less than 0.2 eV/{\AA}, which correspond to the VBM and CBM around $k_0=0$ in the k-space tight-binding method. The counter lines show the phases deviating from the exact 12-fold symmetry (occupation number is 0.5) by 0.05, 0.1, 0.15 and 0.2.}
\label{fig:occup_PE}
\end{figure*}

Since the appearance of the 12-fold symmetric states is attributed to the 12-wave interaction and their degeneration in energy, the 12-fold symmetry of the eigen-states should be unstable if the energy degeneration of the 12 waves is broken. In order to understand this effect clearly, we adopt the k-space tight-binding method proposed by P. Moon et al. \cite{30tbg_Moon} and an electric field is applied perpendicular to the graphene plane to break the energy degeneration of two layers. In this method, a $k_0$ related subspace is spanned by Bloch basis functions of top layer $\lbrace\left|\bm{k}_0 + \bm{G}, \widetilde{X}\right>\rbrace$ and Bloch basis functions of bottom layer $\lbrace\left|\bm{k}_0 + \bm{\widetilde{G}}, X\right>\rbrace$, where $\bm{G}$ ($X$) and $\bm{\widetilde{G}}$ ($\widetilde{X}$) are the reciprocal lattice vectors (sublattice $A$ or $B$) of bottom and top layers, respectively. Any two basis functions with one and another from bottom and top layers, respectively, satisfy the relationship $\braket{\bm{k}_0+\bm{\widetilde{G}}, X|U|\bm{k}_0+\bm{G}, \widetilde{X}}=T(\bm{k}_0+\bm{G}+\bm{\widetilde{G}})e^{-i\bm{\widetilde{G}}\cdot\bm{\tau}_{\widetilde{X}}}e^{i\bm{G}\cdot{\bm{\tau}_{X}}}$, where $U$ is the interlayer interaction and $T(\bm{k}_0+\bm{G}+\bm{\widetilde{G}})$ is the Fourier component of the interlayer hopping function at vector $\bm{k}_0+\bm{G}+\bm{\widetilde{G}}$. In this paper, we only focus on the case around $\bm{k}_0=0$ because quasicrystal electronic states exist only at $\bm{k}_0=0$ exactly. Besides, the 12-wave approximation is used because it has been proven to be valid enough for simulating the electronic properties of 30{\degree} twisted bilayer graphene \cite{30tbg_Moon}. It is worth noting that by comparing the quasi-band structure in the k-space tight-binding method around $\bm{k}_0=0$ with 12-wave approximation and the effective band structure of the 15/26 approximant around $Q$ point, they are very good in agreement with each other. It means around $\bm{k}_0=0$, the 12-wave is exactly enough and increasing the basis size to 182-wave and so on will be over-complete and introduce some redundancy bands, which can not be detected by ARPES measurement. 

\begin{figure}[!htbp]
\centering
\includegraphics[width=6 cm]{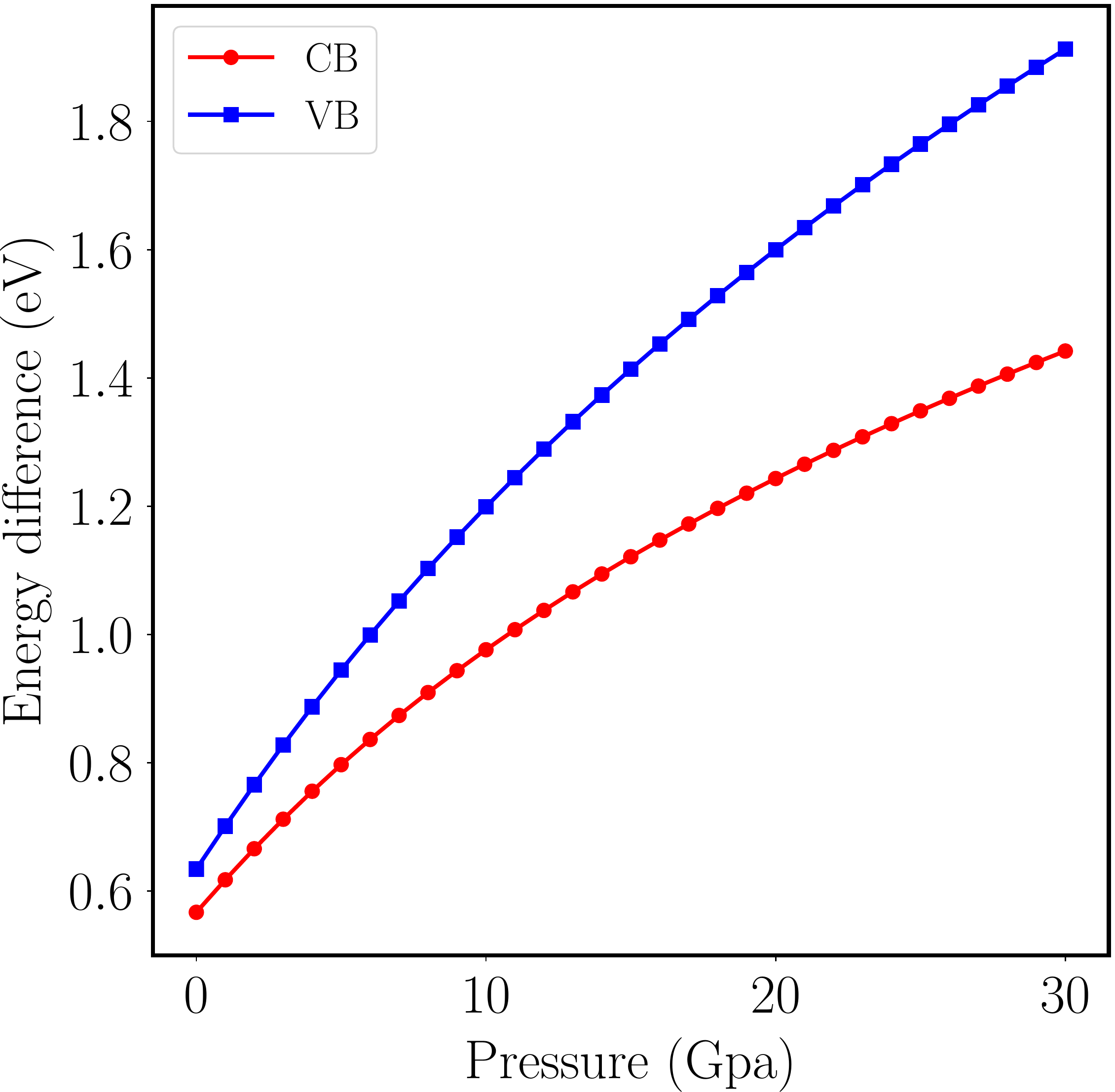}
\caption{The energy differences between bonding and anti-bonding states at zero electric field for the valence band and the conduction band. }
\label{fig:energy_split}
\end{figure}

At $\bm{k}_0=0$ and under the 12-wave approximation, after folding the $\bm{k}$ points into the BZs of the two layers, the subspace is just spanned by the Bloch basis functions $\{\left| Q_i, X\right>\}$ of the bottom layer and the Bloch basis functions $\{\left| \widetilde{Q}_i, \widetilde{X}\right>\}$ of the top layers with $i=0,1,2,3,4,5$ (see Fig. \ref{fig:struct}(b) for their positions). By analysing the eigen-vectors of the 12-wave Hamiltonian at any pressure and electric field, a hybridization picture shown in Fig. \ref{fig:hybrid} can be constructed. The VBM and CBM (labelled by $\left|\delta\right>$ and $\left|\chi\right>$) are the anti-bonding and bonding states after the hybridization between $\left|\alpha \right>$ and $\left|\beta \right>$ states and hybridization between $\left|\lambda\right>$ and $\left|\mu\right>$, respectively. The hybridizations can be expressed by
\begin{equation}
\begin{split}
&\left|\gamma\right> = C_{\gamma\alpha}\left| \alpha \right> + C_{\gamma \beta} \left| \beta\right>\\
&\left|\delta\right> = C_{\delta\alpha}\left| \alpha \right> + C_{\delta \beta} \left| \beta\right>
\end{split}
\end{equation}
and
\begin{equation}
\begin{split}
&\left|\chi\right> = C_{\chi\lambda}\left| \lambda \right> + C_{\chi \mu} \left| \mu\right>\\
&\left|\kappa\right> = C_{\kappa\lambda}\left| \lambda \right> + C_{\kappa \mu} \left| \mu\right>.
\end{split}
\end{equation}
$\left| \alpha \right>$ and $\left| \lambda \right>$ ($\left| \beta \right>$ and $\left| \mu \right>$) exist in the bottom (top) layer.            
These states can be combined by the Bloch basis functions as:
\begin{equation}
\begin{split}
&\left| \alpha \right> =  {{1}\over{\sqrt{6}}} \left( \left|Q_0^- \right>-\left|Q_1^- \right>+\left|Q_2^- \right> -\left|Q_3^- \right>+\left|Q_4^- \right> - \left|Q_5^- \right> \right)\\
&\left| \beta \right> = {{1}\over{\sqrt{6}}} \left( \left|\widetilde{Q}_0^- \right>-\left|\widetilde{Q}_1^- \right>+\left|\widetilde{Q}_2^- \right> -\left|\widetilde{Q}_3^- \right>+\left|\widetilde{Q}_4^- \right> - \left|\widetilde{Q}_5^- \right> \right),
\end{split}
\end{equation}
where,
\begin{equation}
\begin{split}
\left|Q_i^- \right>& ={{1}\over{\sqrt{2}}} \left(\left|Q_i,A \right> - \left|Q_i, B \right>\right) \\
\left|\widetilde{Q}_i^- \right> &={{1}\over{\sqrt{2}}} \left(\left|\widetilde{Q}_i,A \right> - \left|\widetilde{Q}_i, B \right>\right),
\end{split}
\end{equation}
and 
\begin{equation}
\begin{split}
\left| \lambda \right> = & {{1}\over{\sqrt{6}}} \left( \left|Q_0^+ \right>+\left|Q_1^+ \right>+\left|Q_2^+ \right> +\left|Q_3^+ \right>+\left|Q_4^+ \right> + \left|Q_5^+ \right> \right)\\
\left| \mu \right> = & {{1}\over{\sqrt{6}}} \left( \left|\widetilde{Q}_0^+ \right>+\left|\widetilde{Q}_1^+ \right>+\left|\widetilde{Q}_2^+ \right> +\left|\widetilde{Q}_3^+ \right>+\left|\widetilde{Q}_4^+ \right> + \left|\widetilde{Q}_5^+ \right> \right),
\end{split}
\end{equation}
where,
\begin{equation}
\begin{split}
\left|Q_i^+ \right> =&{{1}\over{\sqrt{2}}} \left(\left|Q_i,A \right> + \left|Q_i, B \right>\right) \\
\left|\widetilde{Q}_i^+ \right> =&{{1}\over{\sqrt{2}}} \left(\left|\widetilde{Q}_i,A \right> + \left|\widetilde{Q}_i, B \right>\right).
\end{split}
\end{equation}
The spatial distributions of these states are shown in Fig. \ref{fig:hybrid}. When the electric field is zero, the energy degeneration of the 12 waves is kept, then $\left|C_{\gamma,\alpha} \right|^2=\left|C_{\gamma,\beta} \right|^2=\left|C_{\delta,\alpha} \right|^2=\left|C_{\delta,\beta} \right|^2=\left|C_{\chi,\lambda} \right|^2=\left|C_{\chi,\mu} \right|^2=\left|C_{\kappa,\lambda} \right|^2=\left|C_{\kappa,\mu} \right|^2=0.5$, and all the states $\left| \gamma\right>$, $\left| \delta\right>$, $\left| \chi\right>$ and $\left| \kappa \right>$ are 12-fold symmetric. Besides, as the pressure increases, the hybridization becomes increasingly stronger, which pushes the states $\left| \delta \right>$ and $\left| \chi \right>$ towards the Fermi level gradually. After applying the electric field, the 12-fold symmetry of these hybridized states will be destroyed. As an example, the hybridization picture for the case of 5 GPa pressure and 0.1 eV/{\AA} electric field is shown in Fig. \ref{fig:hybrid}. It shows that the non-equivalent occupation numbers on bottom and top layers makes the hybridized states derivate from the 12-fold symmetry. The evolutions of the occupation numbers of the VBM and CBM on the bottom layer, namely $\left|C_{\delta\alpha}\right|^2$ and $\left|C_{\chi\lambda}\right|^2$, are shown in Fig. \ref{fig:occup_PE}. Two conclusions can be found out from the results. One is that at a specific electric field, the larger pressure can make the occupation number of both VBM and CBM on the bottom layer closer to 0.5, which means that applying pressure is a valid way to recover the 12-fold symmetry of both the VBM and CBM to some extent against the electric field. The reason is that for a specific electric field a higher pressure can results in the smaller interlayer spacing, namely not only the stronger interlayer coupling but also the smaller on-site energy difference between the two layers. Another conclusion is that the 12-fold symmetry of the CBM is easier to be broken than that of the VBM. That is because for a specific pressure, the hybridization strength for the valence band is always stronger than the conduction band. Their hybridization strengths can be described by the energy differences between bonding and anti-bonding states, namely between $\left| \gamma \right>$ and $\left| \delta\right>$ for the valence band and between $\left| \chi \right>$ and $\left| \kappa \right>$ for the conduction band (see Fig. \ref{fig:energy_split}).  

\section{Conclusions}
By using the tight-binding model, we study the dependence of the quasicrystalline electronic states on the external pressure and electric field. We confirm that the pressure can push the energies of these 12-fold symmetric states towards the Fermi level. Such a phenomenon is attributed to the stronger hybridization between the 12 waves of the two layers for higher pressure. Furthermore, the electron or hole doping around $4\times10^{14}\; cm^{-2}$ can tunes the Fermi level to meet these quasicrystalline electronic states exactly, which will make 30{\degree} twisted bilayer graphene manifest the quasicrystalline characters in electronic properties. Moreover, the electric field perpendicular to graphene plane will destroy the 12-fold symmetry of these states. Comparing with the 12-fold symmetric state in the valence band, the 12-fold symmetry of the state in the conduction band is easier to be destroyed. This is because of the stronger hybridization in the valence band than that in the conduction band. Moreover, applying the external pressure can recover the 12-fold symmetry of these states to some extent against the electric field by increasing the interlayer interaction and reducing the on-site energy difference between two layers.

\section*{ACKNOWLEDGEMENTS}
This work is supported by the National Science Foundation of China (Grant No. 11774269) and China Postdoctoral Science Foundation (Grant No. 2018M632902). MIK acknowledges a support by the JTC-FLAGERA Project GRANSPORT. Numerical calculations presented in this paper have been partially performed on the supercomputing system in the Supercomputing Center of Wuhan University. Support by the Netherlands National Computing Facilities foundation (NCF), with funding from the Netherlands Organisation for Scientific Research (NWO), is gratefully acknowledged.

\bibliography{30tBG_pressure}

\begin{thebibliography}{35}%
\makeatletter
\providecommand \@ifxundefined [1]{%
 \@ifx{#1\undefined}
}%
\providecommand \@ifnum [1]{%
 \ifnum #1\expandafter \@firstoftwo
 \else \expandafter \@secondoftwo
 \fi
}%
\providecommand \@ifx [1]{%
 \ifx #1\expandafter \@firstoftwo
 \else \expandafter \@secondoftwo
 \fi
}%
\providecommand \natexlab [1]{#1}%
\providecommand \enquote  [1]{``#1''}%
\providecommand \bibnamefont  [1]{#1}%
\providecommand \bibfnamefont [1]{#1}%
\providecommand \citenamefont [1]{#1}%
\providecommand \href@noop [0]{\@secondoftwo}%
\providecommand \href [0]{\begingroup \@sanitize@url \@href}%
\providecommand \@href[1]{\@@startlink{#1}\@@href}%
\providecommand \@@href[1]{\endgroup#1\@@endlink}%
\providecommand \@sanitize@url [0]{\catcode `\\12\catcode `\$12\catcode
  `\&12\catcode `\#12\catcode `\^12\catcode `\_12\catcode `\%12\relax}%
\providecommand \@@startlink[1]{}%
\providecommand \@@endlink[0]{}%
\providecommand \url  [0]{\begingroup\@sanitize@url \@url }%
\providecommand \@url [1]{\endgroup\@href {#1}{\urlprefix }}%
\providecommand \urlprefix  [0]{URL }%
\providecommand \Eprint [0]{\href }%
\providecommand \doibase [0]{http://dx.doi.org/}%
\providecommand \selectlanguage [0]{\@gobble}%
\providecommand \bibinfo  [0]{\@secondoftwo}%
\providecommand \bibfield  [0]{\@secondoftwo}%
\providecommand \translation [1]{[#1]}%
\providecommand \BibitemOpen [0]{}%
\providecommand \bibitemStop [0]{}%
\providecommand \bibitemNoStop [0]{.\EOS\space}%
\providecommand \EOS [0]{\spacefactor3000\relax}%
\providecommand \BibitemShut  [1]{\csname bibitem#1\endcsname}%
\let\auto@bib@innerbib\@empty
%</preamble>
\bibitem [{\citenamefont {Rakhmanov}\ \emph {et~al.}(2012)\citenamefont
  {Rakhmanov}, \citenamefont {Rozhkov}, \citenamefont {Sboychakov},\ and\
  \citenamefont {Nori}}]{BG_AA0}%
  \BibitemOpen
  \bibfield  {author} {\bibinfo {author} {\bibfnamefont {A.~L.}\ \bibnamefont
  {Rakhmanov}}, \bibinfo {author} {\bibfnamefont {A.~V.}\ \bibnamefont
  {Rozhkov}}, \bibinfo {author} {\bibfnamefont {A.~O.}\ \bibnamefont
  {Sboychakov}}, \ and\ \bibinfo {author} {\bibfnamefont {F.}~\bibnamefont
  {Nori}},\ }\href {\doibase 10.1103/PhysRevLett.109.206801} {\bibfield
  {journal} {\bibinfo  {journal} {Phys. Rev. Lett.}\ }\textbf {\bibinfo
  {volume} {109}},\ \bibinfo {pages} {206801} (\bibinfo {year}
  {2012})}\BibitemShut {NoStop}%
\bibitem [{\citenamefont {Novoselov}\ \emph {et~al.}(2006)\citenamefont
  {Novoselov}, \citenamefont {McCann}, \citenamefont {Morozov}, \citenamefont
  {Fal'ko}, \citenamefont {Katsnelson}, \citenamefont {Zeitler}, \citenamefont
  {Jiang}, \citenamefont {Schedin},\ and\ \citenamefont {Geim}}]{AB_orig0}%
  \BibitemOpen
  \bibfield  {author} {\bibinfo {author} {\bibfnamefont {K.~S.}\ \bibnamefont
  {Novoselov}}, \bibinfo {author} {\bibfnamefont {E.}~\bibnamefont {McCann}},
  \bibinfo {author} {\bibfnamefont {S.~V.}\ \bibnamefont {Morozov}}, \bibinfo
  {author} {\bibfnamefont {V.~I.}\ \bibnamefont {Fal'ko}}, \bibinfo {author}
  {\bibfnamefont {M.~I.}\ \bibnamefont {Katsnelson}}, \bibinfo {author}
  {\bibfnamefont {U.}~\bibnamefont {Zeitler}}, \bibinfo {author} {\bibfnamefont
  {D.}~\bibnamefont {Jiang}}, \bibinfo {author} {\bibfnamefont
  {F.}~\bibnamefont {Schedin}}, \ and\ \bibinfo {author} {\bibfnamefont
  {A.~K.}\ \bibnamefont {Geim}},\ }\href {\doibase 10.1038/nphys245} {\bibfield
   {journal} {\bibinfo  {journal} {Nature Physics}\ }\textbf {\bibinfo {volume}
  {2}},\ \bibinfo {pages} {177} (\bibinfo {year} {2006})}\BibitemShut {NoStop}%
\bibitem [{\citenamefont {McCann}\ and\ \citenamefont
  {Fal'ko}(2006)}]{AB_orig1}%
  \BibitemOpen
  \bibfield  {author} {\bibinfo {author} {\bibfnamefont {E.}~\bibnamefont
  {McCann}}\ and\ \bibinfo {author} {\bibfnamefont {V.~I.}\ \bibnamefont
  {Fal'ko}},\ }\href {\doibase 10.1103/PhysRevLett.96.086805} {\bibfield
  {journal} {\bibinfo  {journal} {Phys. Rev. Lett.}\ }\textbf {\bibinfo
  {volume} {96}},\ \bibinfo {pages} {086805} (\bibinfo {year}
  {2006})}\BibitemShut {NoStop}%
\bibitem [{\citenamefont {Aoki}\ and\ \citenamefont {Amawashi}(2007)}]{BG_AB0}%
  \BibitemOpen
  \bibfield  {author} {\bibinfo {author} {\bibfnamefont {M.}~\bibnamefont
  {Aoki}}\ and\ \bibinfo {author} {\bibfnamefont {H.}~\bibnamefont
  {Amawashi}},\ }\href {\doibase https://doi.org/10.1016/j.ssc.2007.02.013}
  {\bibfield  {journal} {\bibinfo  {journal} {Solid State Communications}\
  }\textbf {\bibinfo {volume} {142}},\ \bibinfo {pages} {123 } (\bibinfo {year}
  {2007})}\BibitemShut {NoStop}%
\bibitem [{\citenamefont {Latil}\ and\ \citenamefont {Henrard}(2006)}]{BG_AB1}%
  \BibitemOpen
  \bibfield  {author} {\bibinfo {author} {\bibfnamefont {S.}~\bibnamefont
  {Latil}}\ and\ \bibinfo {author} {\bibfnamefont {L.}~\bibnamefont
  {Henrard}},\ }\href {\doibase 10.1103/PhysRevLett.97.036803} {\bibfield
  {journal} {\bibinfo  {journal} {Phys. Rev. Lett.}\ }\textbf {\bibinfo
  {volume} {97}},\ \bibinfo {pages} {036803} (\bibinfo {year}
  {2006})}\BibitemShut {NoStop}%
\bibitem [{\citenamefont {Partoens}\ and\ \citenamefont
  {Peeters}(2006)}]{BG_AB2}%
  \BibitemOpen
  \bibfield  {author} {\bibinfo {author} {\bibfnamefont {B.}~\bibnamefont
  {Partoens}}\ and\ \bibinfo {author} {\bibfnamefont {F.~M.}\ \bibnamefont
  {Peeters}},\ }\href {\doibase 10.1103/PhysRevB.74.075404} {\bibfield
  {journal} {\bibinfo  {journal} {Phys. Rev. B}\ }\textbf {\bibinfo {volume}
  {74}},\ \bibinfo {pages} {075404} (\bibinfo {year} {2006})}\BibitemShut
  {NoStop}%
\bibitem [{\citenamefont {Su\'arez~Morell}\ \emph {et~al.}(2010)\citenamefont
  {Su\'arez~Morell}, \citenamefont {Correa}, \citenamefont {Vargas},
  \citenamefont {Pacheco},\ and\ \citenamefont {Barticevic}}]{TBG_flatband}%
  \BibitemOpen
  \bibfield  {author} {\bibinfo {author} {\bibfnamefont {E.}~\bibnamefont
  {Su\'arez~Morell}}, \bibinfo {author} {\bibfnamefont {J.~D.}\ \bibnamefont
  {Correa}}, \bibinfo {author} {\bibfnamefont {P.}~\bibnamefont {Vargas}},
  \bibinfo {author} {\bibfnamefont {M.}~\bibnamefont {Pacheco}}, \ and\
  \bibinfo {author} {\bibfnamefont {Z.}~\bibnamefont {Barticevic}},\ }\href
  {\doibase 10.1103/PhysRevB.82.121407} {\bibfield  {journal} {\bibinfo
  {journal} {Phys. Rev. B}\ }\textbf {\bibinfo {volume} {82}},\ \bibinfo
  {pages} {121407} (\bibinfo {year} {2010})}\BibitemShut {NoStop}%
\bibitem [{\citenamefont {Bistritzer}\ and\ \citenamefont
  {MacDonald}(2011)}]{TBG_flatband1}%
  \BibitemOpen
  \bibfield  {author} {\bibinfo {author} {\bibfnamefont {R.}~\bibnamefont
  {Bistritzer}}\ and\ \bibinfo {author} {\bibfnamefont {A.~H.}\ \bibnamefont
  {MacDonald}},\ }\href {\doibase 10.1073/pnas.1108174108} {\bibfield
  {journal} {\bibinfo  {journal} {Proceedings of the National Academy of
  Sciences}\ }\textbf {\bibinfo {volume} {108}},\ \bibinfo {pages} {12233}
  (\bibinfo {year} {2011})}\BibitemShut {NoStop}%
\bibitem [{\citenamefont {Cao}\ \emph {et~al.}(2018{\natexlab{a}})\citenamefont
  {Cao}, \citenamefont {Fatemi}, \citenamefont {Fang}, \citenamefont
  {Watanabe}, \citenamefont {Taniguchi}, \citenamefont {Kaxiras},\ and\
  \citenamefont {Jarillo-Herrero}}]{BG_superconducting}%
  \BibitemOpen
  \bibfield  {author} {\bibinfo {author} {\bibfnamefont {Y.}~\bibnamefont
  {Cao}}, \bibinfo {author} {\bibfnamefont {V.}~\bibnamefont {Fatemi}},
  \bibinfo {author} {\bibfnamefont {S.}~\bibnamefont {Fang}}, \bibinfo {author}
  {\bibfnamefont {K.}~\bibnamefont {Watanabe}}, \bibinfo {author}
  {\bibfnamefont {T.}~\bibnamefont {Taniguchi}}, \bibinfo {author}
  {\bibfnamefont {E.}~\bibnamefont {Kaxiras}}, \ and\ \bibinfo {author}
  {\bibfnamefont {P.}~\bibnamefont {Jarillo-Herrero}},\ }\href
  {https://doi.org/10.1038/nature26160} {\bibfield  {journal} {\bibinfo
  {journal} {Nature}\ }\textbf {\bibinfo {volume} {556}},\ \bibinfo {pages}
  {43} (\bibinfo {year} {2018}{\natexlab{a}})},\ \bibinfo {note}
  {article}\BibitemShut {NoStop}%
\bibitem [{\citenamefont {Po}\ \emph {et~al.}(2018)\citenamefont {Po},
  \citenamefont {Zou}, \citenamefont {Vishwanath},\ and\ \citenamefont
  {Senthil}}]{TBG_supertivity1}%
  \BibitemOpen
  \bibfield  {author} {\bibinfo {author} {\bibfnamefont {H.~C.}\ \bibnamefont
  {Po}}, \bibinfo {author} {\bibfnamefont {L.}~\bibnamefont {Zou}}, \bibinfo
  {author} {\bibfnamefont {A.}~\bibnamefont {Vishwanath}}, \ and\ \bibinfo
  {author} {\bibfnamefont {T.}~\bibnamefont {Senthil}},\ }\href {\doibase
  10.1103/PhysRevX.8.031089} {\bibfield  {journal} {\bibinfo  {journal} {Phys.
  Rev. X}\ }\textbf {\bibinfo {volume} {8}},\ \bibinfo {pages} {031089}
  (\bibinfo {year} {2018})}\BibitemShut {NoStop}%
\bibitem [{\citenamefont {Yankowitz}\ \emph {et~al.}(2019)\citenamefont
  {Yankowitz}, \citenamefont {Chen}, \citenamefont {Polshyn}, \citenamefont
  {Zhang}, \citenamefont {Watanabe}, \citenamefont {Taniguchi}, \citenamefont
  {Graf}, \citenamefont {Young},\ and\ \citenamefont
  {Dean}}]{TBG_supertivity2}%
  \BibitemOpen
  \bibfield  {author} {\bibinfo {author} {\bibfnamefont {M.}~\bibnamefont
  {Yankowitz}}, \bibinfo {author} {\bibfnamefont {S.}~\bibnamefont {Chen}},
  \bibinfo {author} {\bibfnamefont {H.}~\bibnamefont {Polshyn}}, \bibinfo
  {author} {\bibfnamefont {Y.}~\bibnamefont {Zhang}}, \bibinfo {author}
  {\bibfnamefont {K.}~\bibnamefont {Watanabe}}, \bibinfo {author}
  {\bibfnamefont {T.}~\bibnamefont {Taniguchi}}, \bibinfo {author}
  {\bibfnamefont {D.}~\bibnamefont {Graf}}, \bibinfo {author} {\bibfnamefont
  {A.~F.}\ \bibnamefont {Young}}, \ and\ \bibinfo {author} {\bibfnamefont
  {C.~R.}\ \bibnamefont {Dean}},\ }\href {\doibase 10.1126/science.aav1910}
  {\bibfield  {journal} {\bibinfo  {journal} {Science}\ }\textbf {\bibinfo
  {volume} {363}},\ \bibinfo {pages} {1059} (\bibinfo {year}
  {2019})}\BibitemShut {NoStop}%
\bibitem [{\citenamefont {Cao}\ \emph {et~al.}(2018{\natexlab{b}})\citenamefont
  {Cao}, \citenamefont {Fatemi}, \citenamefont {Demir}, \citenamefont {Fang},
  \citenamefont {Tomarken}, \citenamefont {Luo}, \citenamefont
  {Sanchez-Yamagishi}, \citenamefont {Watanabe}, \citenamefont {Taniguchi},
  \citenamefont {Kaxiras}, \citenamefont {Ashoori},\ and\ \citenamefont
  {Jarillo-Herrero}}]{TBG_insulator_phase}%
  \BibitemOpen
  \bibfield  {author} {\bibinfo {author} {\bibfnamefont {Y.}~\bibnamefont
  {Cao}}, \bibinfo {author} {\bibfnamefont {V.}~\bibnamefont {Fatemi}},
  \bibinfo {author} {\bibfnamefont {A.}~\bibnamefont {Demir}}, \bibinfo
  {author} {\bibfnamefont {S.}~\bibnamefont {Fang}}, \bibinfo {author}
  {\bibfnamefont {S.~L.}\ \bibnamefont {Tomarken}}, \bibinfo {author}
  {\bibfnamefont {J.~Y.}\ \bibnamefont {Luo}}, \bibinfo {author} {\bibfnamefont
  {J.~D.}\ \bibnamefont {Sanchez-Yamagishi}}, \bibinfo {author} {\bibfnamefont
  {K.}~\bibnamefont {Watanabe}}, \bibinfo {author} {\bibfnamefont
  {T.}~\bibnamefont {Taniguchi}}, \bibinfo {author} {\bibfnamefont
  {E.}~\bibnamefont {Kaxiras}}, \bibinfo {author} {\bibfnamefont {R.~C.}\
  \bibnamefont {Ashoori}}, \ and\ \bibinfo {author} {\bibfnamefont
  {P.}~\bibnamefont {Jarillo-Herrero}},\ }\href
  {https://doi.org/10.1038/nature26154} {\bibfield  {journal} {\bibinfo
  {journal} {Nature}\ }\textbf {\bibinfo {volume} {556}},\ \bibinfo {pages}
  {80} (\bibinfo {year} {2018}{\natexlab{b}})}\BibitemShut {NoStop}%
\bibitem [{\citenamefont {Trambly~de Laissardi\`ere}\ \emph
  {et~al.}(2012)\citenamefont {Trambly~de Laissardi\`ere}, \citenamefont
  {Mayou},\ and\ \citenamefont {Magaud}}]{TBG_commensurate_condition}%
  \BibitemOpen
  \bibfield  {author} {\bibinfo {author} {\bibfnamefont {G.}~\bibnamefont
  {Trambly~de Laissardi\`ere}}, \bibinfo {author} {\bibfnamefont
  {D.}~\bibnamefont {Mayou}}, \ and\ \bibinfo {author} {\bibfnamefont
  {L.}~\bibnamefont {Magaud}},\ }\href {\doibase 10.1103/PhysRevB.86.125413}
  {\bibfield  {journal} {\bibinfo  {journal} {Phys. Rev. B}\ }\textbf {\bibinfo
  {volume} {86}},\ \bibinfo {pages} {125413} (\bibinfo {year}
  {2012})}\BibitemShut {NoStop}%
\bibitem [{\citenamefont {Yao}\ \emph {et~al.}(2018)\citenamefont {Yao},
  \citenamefont {Wang}, \citenamefont {Bao}, \citenamefont {Zhang},
  \citenamefont {Zhang}, \citenamefont {Bao}, \citenamefont {Chan},
  \citenamefont {Chen}, \citenamefont {Avila}, \citenamefont {Asensio},
  \citenamefont {Zhu},\ and\ \citenamefont {Zhou}}]{pnas_QC}%
  \BibitemOpen
  \bibfield  {author} {\bibinfo {author} {\bibfnamefont {W.}~\bibnamefont
  {Yao}}, \bibinfo {author} {\bibfnamefont {E.}~\bibnamefont {Wang}}, \bibinfo
  {author} {\bibfnamefont {C.}~\bibnamefont {Bao}}, \bibinfo {author}
  {\bibfnamefont {Y.}~\bibnamefont {Zhang}}, \bibinfo {author} {\bibfnamefont
  {K.}~\bibnamefont {Zhang}}, \bibinfo {author} {\bibfnamefont
  {K.}~\bibnamefont {Bao}}, \bibinfo {author} {\bibfnamefont {C.~K.}\
  \bibnamefont {Chan}}, \bibinfo {author} {\bibfnamefont {C.}~\bibnamefont
  {Chen}}, \bibinfo {author} {\bibfnamefont {J.}~\bibnamefont {Avila}},
  \bibinfo {author} {\bibfnamefont {M.~C.}\ \bibnamefont {Asensio}}, \bibinfo
  {author} {\bibfnamefont {J.}~\bibnamefont {Zhu}}, \ and\ \bibinfo {author}
  {\bibfnamefont {S.}~\bibnamefont {Zhou}},\ }\href {\doibase
  10.1073/pnas.1720865115} {\bibfield  {journal} {\bibinfo  {journal}
  {Proceedings of the National Academy of Sciences}\ }\textbf {\bibinfo
  {volume} {115}},\ \bibinfo {pages} {6928} (\bibinfo {year}
  {2018})}\BibitemShut {NoStop}%
\bibitem [{\citenamefont {Ahn}\ \emph {et~al.}(2018)\citenamefont {Ahn},
  \citenamefont {Moon}, \citenamefont {Kim}, \citenamefont {Kim}, \citenamefont
  {Shin}, \citenamefont {Kim}, \citenamefont {Cha}, \citenamefont {Kahng},
  \citenamefont {Kim}, \citenamefont {Koshino}, \citenamefont {Son},
  \citenamefont {Yang},\ and\ \citenamefont {Ahn}}]{science_QC}%
  \BibitemOpen
  \bibfield  {author} {\bibinfo {author} {\bibfnamefont {S.~J.}\ \bibnamefont
  {Ahn}}, \bibinfo {author} {\bibfnamefont {P.}~\bibnamefont {Moon}}, \bibinfo
  {author} {\bibfnamefont {T.-H.}\ \bibnamefont {Kim}}, \bibinfo {author}
  {\bibfnamefont {H.-W.}\ \bibnamefont {Kim}}, \bibinfo {author} {\bibfnamefont
  {H.-C.}\ \bibnamefont {Shin}}, \bibinfo {author} {\bibfnamefont {E.~H.}\
  \bibnamefont {Kim}}, \bibinfo {author} {\bibfnamefont {H.~W.}\ \bibnamefont
  {Cha}}, \bibinfo {author} {\bibfnamefont {S.-J.}\ \bibnamefont {Kahng}},
  \bibinfo {author} {\bibfnamefont {P.}~\bibnamefont {Kim}}, \bibinfo {author}
  {\bibfnamefont {M.}~\bibnamefont {Koshino}}, \bibinfo {author} {\bibfnamefont
  {Y.-W.}\ \bibnamefont {Son}}, \bibinfo {author} {\bibfnamefont {C.-W.}\
  \bibnamefont {Yang}}, \ and\ \bibinfo {author} {\bibfnamefont {J.~R.}\
  \bibnamefont {Ahn}},\ }\href {\doibase 10.1126/science.aar8412} {\bibfield
  {journal} {\bibinfo  {journal} {Science}\ }\textbf {\bibinfo {volume}
  {361}},\ \bibinfo {pages} {782} (\bibinfo {year} {2018})}\BibitemShut
  {NoStop}%
\bibitem [{\citenamefont {Takesaki}\ \emph {et~al.}(2016)\citenamefont
  {Takesaki}, \citenamefont {Kawahara}, \citenamefont {Hibino}, \citenamefont
  {Okada}, \citenamefont {Tsuji},\ and\ \citenamefont {Ago}}]{cm_QC}%
  \BibitemOpen
  \bibfield  {author} {\bibinfo {author} {\bibfnamefont {Y.}~\bibnamefont
  {Takesaki}}, \bibinfo {author} {\bibfnamefont {K.}~\bibnamefont {Kawahara}},
  \bibinfo {author} {\bibfnamefont {H.}~\bibnamefont {Hibino}}, \bibinfo
  {author} {\bibfnamefont {S.}~\bibnamefont {Okada}}, \bibinfo {author}
  {\bibfnamefont {M.}~\bibnamefont {Tsuji}}, \ and\ \bibinfo {author}
  {\bibfnamefont {H.}~\bibnamefont {Ago}},\ }\href {\doibase
  10.1021/acs.chemmater.6b01137} {\bibfield  {journal} {\bibinfo  {journal}
  {Chemistry of Materials}\ }\textbf {\bibinfo {volume} {28}},\ \bibinfo
  {pages} {4583} (\bibinfo {year} {2016})}\BibitemShut {NoStop}%
\bibitem [{\citenamefont {Yan}\ \emph {et~al.}(2019)\citenamefont {Yan},
  \citenamefont {Ma}, \citenamefont {Qiao}, \citenamefont {Zhong},
  \citenamefont {Yang}, \citenamefont {Li}, \citenamefont {Fu}, \citenamefont
  {Zhang},\ and\ \citenamefont {He}}]{30TBG_STM}%
  \BibitemOpen
  \bibfield  {author} {\bibinfo {author} {\bibfnamefont {C.}~\bibnamefont
  {Yan}}, \bibinfo {author} {\bibfnamefont {D.-L.}\ \bibnamefont {Ma}},
  \bibinfo {author} {\bibfnamefont {J.-B.}\ \bibnamefont {Qiao}}, \bibinfo
  {author} {\bibfnamefont {H.-Y.}\ \bibnamefont {Zhong}}, \bibinfo {author}
  {\bibfnamefont {L.}~\bibnamefont {Yang}}, \bibinfo {author} {\bibfnamefont
  {S.-Y.}\ \bibnamefont {Li}}, \bibinfo {author} {\bibfnamefont {Z.-Q.}\
  \bibnamefont {Fu}}, \bibinfo {author} {\bibfnamefont {Y.}~\bibnamefont
  {Zhang}}, \ and\ \bibinfo {author} {\bibfnamefont {L.}~\bibnamefont {He}},\
  }\href {\doibase 10.1088/2053-1583/ab3b16} {\bibfield  {journal} {\bibinfo
  {journal} {2D Materials}\ }\textbf {\bibinfo {volume} {6}},\ \bibinfo {pages}
  {045041} (\bibinfo {year} {2019})}\BibitemShut {NoStop}%
\bibitem [{\citenamefont {{Lin}}\ \emph {et~al.}(2018)\citenamefont {{Lin}},
  \citenamefont {{Samiseresht}}, \citenamefont {{Franke}}, \citenamefont
  {{Parhizkar}}, \citenamefont {{Soubatch}}, \citenamefont {{Amorim}},
  \citenamefont {{Lee}}, \citenamefont {{Kumpf}}, \citenamefont {{Tautz}},\
  and\ \citenamefont {{Bocquet}}}]{30TBG_SiC_arXiv}%
  \BibitemOpen
  \bibfield  {author} {\bibinfo {author} {\bibfnamefont {Y.~R.}\ \bibnamefont
  {{Lin}}}, \bibinfo {author} {\bibfnamefont {N.}~\bibnamefont
  {{Samiseresht}}}, \bibinfo {author} {\bibfnamefont {M.}~\bibnamefont
  {{Franke}}}, \bibinfo {author} {\bibfnamefont {S.}~\bibnamefont
  {{Parhizkar}}}, \bibinfo {author} {\bibfnamefont {S.}~\bibnamefont
  {{Soubatch}}}, \bibinfo {author} {\bibfnamefont {B.}~\bibnamefont
  {{Amorim}}}, \bibinfo {author} {\bibfnamefont {T.~L.}\ \bibnamefont {{Lee}}},
  \bibinfo {author} {\bibfnamefont {C.}~\bibnamefont {{Kumpf}}}, \bibinfo
  {author} {\bibfnamefont {F.~S.}\ \bibnamefont {{Tautz}}}, \ and\ \bibinfo
  {author} {\bibfnamefont {F.~C.}\ \bibnamefont {{Bocquet}}},\ }\href
  {https://arxiv.org/abs/1809.07958} {\bibfield  {journal} {\bibinfo  {journal}
  {arXiv e-prints}\ ,\ \bibinfo {pages} {arXiv:1809.07958}} (\bibinfo {year}
  {2018})}\BibitemShut {NoStop}%
\bibitem [{\citenamefont {Deng}\ \emph {et~al.}(2020)\citenamefont {Deng},
  \citenamefont {Wang}, \citenamefont {Li}, \citenamefont {Li}, \citenamefont
  {Wang}, \citenamefont {Tang}, \citenamefont {Fu}, \citenamefont {Tian},
  \citenamefont {Gao}, \citenamefont {Xue},\ and\ \citenamefont
  {Peng}}]{30tBG_on_Cu_ACSNano}%
  \BibitemOpen
  \bibfield  {author} {\bibinfo {author} {\bibfnamefont {B.}~\bibnamefont
  {Deng}}, \bibinfo {author} {\bibfnamefont {B.}~\bibnamefont {Wang}}, \bibinfo
  {author} {\bibfnamefont {N.}~\bibnamefont {Li}}, \bibinfo {author}
  {\bibfnamefont {R.}~\bibnamefont {Li}}, \bibinfo {author} {\bibfnamefont
  {Y.}~\bibnamefont {Wang}}, \bibinfo {author} {\bibfnamefont {J.}~\bibnamefont
  {Tang}}, \bibinfo {author} {\bibfnamefont {Q.}~\bibnamefont {Fu}}, \bibinfo
  {author} {\bibfnamefont {Z.}~\bibnamefont {Tian}}, \bibinfo {author}
  {\bibfnamefont {P.}~\bibnamefont {Gao}}, \bibinfo {author} {\bibfnamefont
  {J.}~\bibnamefont {Xue}}, \ and\ \bibinfo {author} {\bibfnamefont
  {H.}~\bibnamefont {Peng}},\ }\href {\doibase 10.1021/acsnano.9b07091}
  {\bibfield  {journal} {\bibinfo  {journal} {ACS Nano}\ }\textbf {\bibinfo
  {volume} {14}},\ \bibinfo {pages} {1656} (\bibinfo {year}
  {2020})}\BibitemShut {NoStop}%
\bibitem [{\citenamefont {{Pezzini}}\ \emph {et~al.}(2020)\citenamefont
  {{Pezzini}}, \citenamefont {{Miseikis}}, \citenamefont {{Piccinini}},
  \citenamefont {{Forti}}, \citenamefont {{Pace}}, \citenamefont {{Engelke}},
  \citenamefont {{Rossella}}, \citenamefont {{Watanabe}}, \citenamefont
  {{Taniguchi}}, \citenamefont {{Kim}},\ and\ \citenamefont
  {{Coletti}}}]{30tBG_onCu_arXiv}%
  \BibitemOpen
  \bibfield  {author} {\bibinfo {author} {\bibfnamefont {S.}~\bibnamefont
  {{Pezzini}}}, \bibinfo {author} {\bibfnamefont {V.}~\bibnamefont
  {{Miseikis}}}, \bibinfo {author} {\bibfnamefont {G.}~\bibnamefont
  {{Piccinini}}}, \bibinfo {author} {\bibfnamefont {S.}~\bibnamefont
  {{Forti}}}, \bibinfo {author} {\bibfnamefont {S.}~\bibnamefont {{Pace}}},
  \bibinfo {author} {\bibfnamefont {R.}~\bibnamefont {{Engelke}}}, \bibinfo
  {author} {\bibfnamefont {F.}~\bibnamefont {{Rossella}}}, \bibinfo {author}
  {\bibfnamefont {K.}~\bibnamefont {{Watanabe}}}, \bibinfo {author}
  {\bibfnamefont {T.}~\bibnamefont {{Taniguchi}}}, \bibinfo {author}
  {\bibfnamefont {P.}~\bibnamefont {{Kim}}}, \ and\ \bibinfo {author}
  {\bibfnamefont {C.}~\bibnamefont {{Coletti}}},\ }\href@noop {} {\bibfield
  {journal} {\bibinfo  {journal} {arXiv e-prints}\ ,\ \bibinfo {eid}
  {arXiv:2001.10427}} (\bibinfo {year} {2020})},\ \Eprint
  {http://arxiv.org/abs/2001.10427} {arXiv:2001.10427 [cond-mat.mes-hall]}
  \BibitemShut {NoStop}%
\bibitem [{\citenamefont {Suzuki}\ \emph {et~al.}(2019)\citenamefont {Suzuki},
  \citenamefont {Iimori}, \citenamefont {Ahn}, \citenamefont {Zhao},
  \citenamefont {Watanabe}, \citenamefont {Xu}, \citenamefont {Fujisawa},
  \citenamefont {Kanai}, \citenamefont {Ishii}, \citenamefont {Itatani},
  \citenamefont {Suwa}, \citenamefont {Fukidome}, \citenamefont {Tanaka},
  \citenamefont {Ahn}, \citenamefont {Okazaki}, \citenamefont {Shin},
  \citenamefont {Komori},\ and\ \citenamefont
  {Matsuda}}]{30TBG_Ultrafast_unbalance_arXiv}%
  \BibitemOpen
  \bibfield  {author} {\bibinfo {author} {\bibfnamefont {T.}~\bibnamefont
  {Suzuki}}, \bibinfo {author} {\bibfnamefont {T.}~\bibnamefont {Iimori}},
  \bibinfo {author} {\bibfnamefont {S.~J.}\ \bibnamefont {Ahn}}, \bibinfo
  {author} {\bibfnamefont {Y.}~\bibnamefont {Zhao}}, \bibinfo {author}
  {\bibfnamefont {M.}~\bibnamefont {Watanabe}}, \bibinfo {author}
  {\bibfnamefont {J.}~\bibnamefont {Xu}}, \bibinfo {author} {\bibfnamefont
  {M.}~\bibnamefont {Fujisawa}}, \bibinfo {author} {\bibfnamefont
  {T.}~\bibnamefont {Kanai}}, \bibinfo {author} {\bibfnamefont
  {N.}~\bibnamefont {Ishii}}, \bibinfo {author} {\bibfnamefont
  {J.}~\bibnamefont {Itatani}}, \bibinfo {author} {\bibfnamefont
  {K.}~\bibnamefont {Suwa}}, \bibinfo {author} {\bibfnamefont {H.}~\bibnamefont
  {Fukidome}}, \bibinfo {author} {\bibfnamefont {S.}~\bibnamefont {Tanaka}},
  \bibinfo {author} {\bibfnamefont {J.~R.}\ \bibnamefont {Ahn}}, \bibinfo
  {author} {\bibfnamefont {K.}~\bibnamefont {Okazaki}}, \bibinfo {author}
  {\bibfnamefont {S.}~\bibnamefont {Shin}}, \bibinfo {author} {\bibfnamefont
  {F.}~\bibnamefont {Komori}}, \ and\ \bibinfo {author} {\bibfnamefont
  {I.}~\bibnamefont {Matsuda}},\ }\href {\doibase 10.1021/acsnano.9b06091}
  {\bibfield  {journal} {\bibinfo  {journal} {ACS Nano}\ }\textbf {\bibinfo
  {volume} {13}},\ \bibinfo {pages} {11981} (\bibinfo {year}
  {2019})}\BibitemShut {NoStop}%
\bibitem [{\citenamefont {Park}\ \emph {et~al.}(2019)\citenamefont {Park},
  \citenamefont {Kim},\ and\ \citenamefont {Lee}}]{30TBG_localization}%
  \BibitemOpen
  \bibfield  {author} {\bibinfo {author} {\bibfnamefont {M.~J.}\ \bibnamefont
  {Park}}, \bibinfo {author} {\bibfnamefont {H.~S.}\ \bibnamefont {Kim}}, \
  and\ \bibinfo {author} {\bibfnamefont {S.}~\bibnamefont {Lee}},\ }\href
  {\doibase 10.1103/PhysRevB.99.245401} {\bibfield  {journal} {\bibinfo
  {journal} {Phys. Rev. B}\ }\textbf {\bibinfo {volume} {99}},\ \bibinfo
  {pages} {245401} (\bibinfo {year} {2019})}\BibitemShut {NoStop}%
\bibitem [{\citenamefont {Spurrier}\ and\ \citenamefont
  {Cooper}(2019)}]{30TBG_quantum_oscillation}%
  \BibitemOpen
  \bibfield  {author} {\bibinfo {author} {\bibfnamefont {S.}~\bibnamefont
  {Spurrier}}\ and\ \bibinfo {author} {\bibfnamefont {N.~R.}\ \bibnamefont
  {Cooper}},\ }\href {\doibase 10.1103/PhysRevB.100.081405} {\bibfield
  {journal} {\bibinfo  {journal} {Phys. Rev. B}\ }\textbf {\bibinfo {volume}
  {100}},\ \bibinfo {pages} {081405} (\bibinfo {year} {2019})}\BibitemShut
  {NoStop}%
\bibitem [{\citenamefont {Koren}\ and\ \citenamefont
  {Duerig}(2016)}]{30TBG_superlubricity}%
  \BibitemOpen
  \bibfield  {author} {\bibinfo {author} {\bibfnamefont {E.}~\bibnamefont
  {Koren}}\ and\ \bibinfo {author} {\bibfnamefont {U.}~\bibnamefont {Duerig}},\
  }\href {\doibase 10.1103/PhysRevB.93.201404} {\bibfield  {journal} {\bibinfo
  {journal} {Phys. Rev. B}\ }\textbf {\bibinfo {volume} {93}},\ \bibinfo
  {pages} {201404} (\bibinfo {year} {2016})}\BibitemShut {NoStop}%
\bibitem [{\citenamefont {Moon}\ \emph {et~al.}(2019)\citenamefont {Moon},
  \citenamefont {Koshino},\ and\ \citenamefont {Son}}]{30tbg_Moon}%
  \BibitemOpen
  \bibfield  {author} {\bibinfo {author} {\bibfnamefont {P.}~\bibnamefont
  {Moon}}, \bibinfo {author} {\bibfnamefont {M.}~\bibnamefont {Koshino}}, \
  and\ \bibinfo {author} {\bibfnamefont {Y.-W.}\ \bibnamefont {Son}},\ }\href
  {\doibase 10.1103/PhysRevB.99.165430} {\bibfield  {journal} {\bibinfo
  {journal} {Phys. Rev. B}\ }\textbf {\bibinfo {volume} {99}},\ \bibinfo
  {pages} {165430} (\bibinfo {year} {2019})}\BibitemShut {NoStop}%
\bibitem [{\citenamefont {Carr}\ \emph {et~al.}(2018)\citenamefont {Carr},
  \citenamefont {Fang}, \citenamefont {Jarillo-Herrero},\ and\ \citenamefont
  {Kaxiras}}]{TB_wannier_pressure}%
  \BibitemOpen
  \bibfield  {author} {\bibinfo {author} {\bibfnamefont {S.}~\bibnamefont
  {Carr}}, \bibinfo {author} {\bibfnamefont {S.}~\bibnamefont {Fang}}, \bibinfo
  {author} {\bibfnamefont {P.}~\bibnamefont {Jarillo-Herrero}}, \ and\ \bibinfo
  {author} {\bibfnamefont {E.}~\bibnamefont {Kaxiras}},\ }\href {\doibase
  10.1103/PhysRevB.98.085144} {\bibfield  {journal} {\bibinfo  {journal} {Phys.
  Rev. B}\ }\textbf {\bibinfo {volume} {98}},\ \bibinfo {pages} {085144}
  (\bibinfo {year} {2018})}\BibitemShut {NoStop}%
\bibitem [{\citenamefont {Fang}\ and\ \citenamefont
  {Kaxiras}(2016)}]{TB_wannier}%
  \BibitemOpen
  \bibfield  {author} {\bibinfo {author} {\bibfnamefont {S.}~\bibnamefont
  {Fang}}\ and\ \bibinfo {author} {\bibfnamefont {E.}~\bibnamefont {Kaxiras}},\
  }\href {\doibase 10.1103/PhysRevB.93.235153} {\bibfield  {journal} {\bibinfo
  {journal} {Phys. Rev. B}\ }\textbf {\bibinfo {volume} {93}},\ \bibinfo
  {pages} {235153} (\bibinfo {year} {2016})}\BibitemShut {NoStop}%
\bibitem [{\citenamefont {Chittari}\ \emph {et~al.}(2018)\citenamefont
  {Chittari}, \citenamefont {Leconte}, \citenamefont {Javvaji},\ and\
  \citenamefont {Jung}}]{Murnaghan_equation}%
  \BibitemOpen
  \bibfield  {author} {\bibinfo {author} {\bibfnamefont {B.~L.}\ \bibnamefont
  {Chittari}}, \bibinfo {author} {\bibfnamefont {N.}~\bibnamefont {Leconte}},
  \bibinfo {author} {\bibfnamefont {S.}~\bibnamefont {Javvaji}}, \ and\
  \bibinfo {author} {\bibfnamefont {J.}~\bibnamefont {Jung}},\ }\href {\doibase
  10.1088/2516-1075/aaead3} {\bibfield  {journal} {\bibinfo  {journal}
  {Electronic Structure}\ }\textbf {\bibinfo {volume} {1}},\ \bibinfo {pages}
  {015001} (\bibinfo {year} {2018})}\BibitemShut {NoStop}%
\bibitem [{\citenamefont {Yankowitz}\ \emph {et~al.}(2018)\citenamefont
  {Yankowitz}, \citenamefont {Jung}, \citenamefont {Laksono}, \citenamefont
  {Leconte}, \citenamefont {Chittari}, \citenamefont {Watanabe}, \citenamefont
  {Taniguchi}, \citenamefont {Adam}, \citenamefont {Graf},\ and\ \citenamefont
  {Dean}}]{hBN_pressure}%
  \BibitemOpen
  \bibfield  {author} {\bibinfo {author} {\bibfnamefont {M.}~\bibnamefont
  {Yankowitz}}, \bibinfo {author} {\bibfnamefont {J.}~\bibnamefont {Jung}},
  \bibinfo {author} {\bibfnamefont {E.}~\bibnamefont {Laksono}}, \bibinfo
  {author} {\bibfnamefont {N.}~\bibnamefont {Leconte}}, \bibinfo {author}
  {\bibfnamefont {B.~L.}\ \bibnamefont {Chittari}}, \bibinfo {author}
  {\bibfnamefont {K.}~\bibnamefont {Watanabe}}, \bibinfo {author}
  {\bibfnamefont {T.}~\bibnamefont {Taniguchi}}, \bibinfo {author}
  {\bibfnamefont {S.}~\bibnamefont {Adam}}, \bibinfo {author} {\bibfnamefont
  {D.}~\bibnamefont {Graf}}, \ and\ \bibinfo {author} {\bibfnamefont {C.~R.}\
  \bibnamefont {Dean}},\ }\href {\doibase 10.1038/s41586-018-0107-1} {\bibfield
   {journal} {\bibinfo  {journal} {Nature}\ }\textbf {\bibinfo {volume}
  {557}},\ \bibinfo {pages} {404} (\bibinfo {year} {2018})}\BibitemShut
  {NoStop}%
\bibitem [{\citenamefont {Yu}\ \emph {et~al.}(2019)\citenamefont {Yu},
  \citenamefont {Wu}, \citenamefont {Zhan}, \citenamefont {Katsnelson},\ and\
  \citenamefont {Yuan}}]{Yu2019}%
  \BibitemOpen
  \bibfield  {author} {\bibinfo {author} {\bibfnamefont {G.}~\bibnamefont
  {Yu}}, \bibinfo {author} {\bibfnamefont {Z.}~\bibnamefont {Wu}}, \bibinfo
  {author} {\bibfnamefont {Z.}~\bibnamefont {Zhan}}, \bibinfo {author}
  {\bibfnamefont {M.~I.}\ \bibnamefont {Katsnelson}}, \ and\ \bibinfo {author}
  {\bibfnamefont {S.}~\bibnamefont {Yuan}},\ }\href {\doibase
  10.1038/s41524-019-0258-0} {\bibfield  {journal} {\bibinfo  {journal} {npj
  Computational Materials}\ }\textbf {\bibinfo {volume} {5}},\ \bibinfo {pages}
  {122} (\bibinfo {year} {2019})}\BibitemShut {NoStop}%
\bibitem [{\citenamefont {{Yu}}\ \emph {et~al.}(2019)\citenamefont {{Yu}},
  \citenamefont {{Wu}}, \citenamefont {{Zhan}}, \citenamefont {{Katsnelson}},\
  and\ \citenamefont {{Yuan}}}]{tMGQ}%
  \BibitemOpen
  \bibfield  {author} {\bibinfo {author} {\bibfnamefont {G.}~\bibnamefont
  {{Yu}}}, \bibinfo {author} {\bibfnamefont {Z.}~\bibnamefont {{Wu}}}, \bibinfo
  {author} {\bibfnamefont {Z.}~\bibnamefont {{Zhan}}}, \bibinfo {author}
  {\bibfnamefont {M.~I.}\ \bibnamefont {{Katsnelson}}}, \ and\ \bibinfo
  {author} {\bibfnamefont {S.}~\bibnamefont {{Yuan}}},\ }\href
  {https://arxiv.org/abs/1908.08439} {\bibfield  {journal} {\bibinfo  {journal}
  {arXiv e-prints}\ ,\ \bibinfo {eid} {arXiv:1908.08439}} (\bibinfo {year}
  {2019})},\ \Eprint {http://arxiv.org/abs/1908.08439} {arXiv:1908.08439
  [cond-mat.mtrl-sci]} \BibitemShut {NoStop}%
\bibitem [{\citenamefont {Nishi}\ \emph {et~al.}(2017)\citenamefont {Nishi},
  \citenamefont {Matsushita},\ and\ \citenamefont {Oshiyama}}]{BandUnfold_SF}%
  \BibitemOpen
  \bibfield  {author} {\bibinfo {author} {\bibfnamefont {H.}~\bibnamefont
  {Nishi}}, \bibinfo {author} {\bibfnamefont {Y.-i.}\ \bibnamefont
  {Matsushita}}, \ and\ \bibinfo {author} {\bibfnamefont {A.}~\bibnamefont
  {Oshiyama}},\ }\href {\doibase 10.1103/PhysRevB.95.085420} {\bibfield
  {journal} {\bibinfo  {journal} {Phys. Rev. B}\ }\textbf {\bibinfo {volume}
  {95}},\ \bibinfo {pages} {085420} (\bibinfo {year} {2017})}\BibitemShut
  {NoStop}%
\bibitem [{\citenamefont {Medeiros}\ \emph {et~al.}(2014)\citenamefont
  {Medeiros}, \citenamefont {Stafstr\"om},\ and\ \citenamefont
  {Bj\"ork}}]{BandUnfold_EBS}%
  \BibitemOpen
  \bibfield  {author} {\bibinfo {author} {\bibfnamefont {P.~V.~C.}\
  \bibnamefont {Medeiros}}, \bibinfo {author} {\bibfnamefont {S.}~\bibnamefont
  {Stafstr\"om}}, \ and\ \bibinfo {author} {\bibfnamefont {J.}~\bibnamefont
  {Bj\"ork}},\ }\href {\doibase 10.1103/PhysRevB.89.041407} {\bibfield
  {journal} {\bibinfo  {journal} {Phys. Rev. B}\ }\textbf {\bibinfo {volume}
  {89}},\ \bibinfo {pages} {041407} (\bibinfo {year} {2014})}\BibitemShut
  {NoStop}%
\bibitem [{\citenamefont {Ye}\ \emph {et~al.}(2012)\citenamefont {Ye},
  \citenamefont {Zhang}, \citenamefont {Akashi}, \citenamefont {Bahramy},
  \citenamefont {Arita},\ and\ \citenamefont {Iwasa}}]{doping0}%
  \BibitemOpen
  \bibfield  {author} {\bibinfo {author} {\bibfnamefont {J.~T.}\ \bibnamefont
  {Ye}}, \bibinfo {author} {\bibfnamefont {Y.~J.}\ \bibnamefont {Zhang}},
  \bibinfo {author} {\bibfnamefont {R.}~\bibnamefont {Akashi}}, \bibinfo
  {author} {\bibfnamefont {M.~S.}\ \bibnamefont {Bahramy}}, \bibinfo {author}
  {\bibfnamefont {R.}~\bibnamefont {Arita}}, \ and\ \bibinfo {author}
  {\bibfnamefont {Y.}~\bibnamefont {Iwasa}},\ }\href {\doibase
  10.1126/science.1228006} {\bibfield  {journal} {\bibinfo  {journal}
  {Science}\ }\textbf {\bibinfo {volume} {338}},\ \bibinfo {pages} {1193}
  (\bibinfo {year} {2012})}\BibitemShut {NoStop}%
\bibitem [{\citenamefont {Piatti}\ \emph {et~al.}(2018)\citenamefont {Piatti},
  \citenamefont {De~Fazio}, \citenamefont {Daghero}, \citenamefont
  {Tamalampudi}, \citenamefont {Yoon}, \citenamefont {Ferrari},\ and\
  \citenamefont {Gonnelli}}]{doping1}%
  \BibitemOpen
  \bibfield  {author} {\bibinfo {author} {\bibfnamefont {E.}~\bibnamefont
  {Piatti}}, \bibinfo {author} {\bibfnamefont {D.}~\bibnamefont {De~Fazio}},
  \bibinfo {author} {\bibfnamefont {D.}~\bibnamefont {Daghero}}, \bibinfo
  {author} {\bibfnamefont {S.~R.}\ \bibnamefont {Tamalampudi}}, \bibinfo
  {author} {\bibfnamefont {D.}~\bibnamefont {Yoon}}, \bibinfo {author}
  {\bibfnamefont {A.~C.}\ \bibnamefont {Ferrari}}, \ and\ \bibinfo {author}
  {\bibfnamefont {R.~S.}\ \bibnamefont {Gonnelli}},\ }\href {\doibase
  10.1021/acs.nanolett.8b01390} {\bibfield  {journal} {\bibinfo  {journal}
  {Nano Letters}\ }\textbf {\bibinfo {volume} {18}},\ \bibinfo {pages} {4821}
  (\bibinfo {year} {2018})}\BibitemShut {NoStop}%
\end{thebibliography}%
\end{document}